\documentclass[10pt,journal,compsoc]{IEEEtran}

\usepackage[nocompress]{cite}
\usepackage{subfigure}
\usepackage{multirow}
\usepackage{graphicx}
\usepackage[space]{grffile}
\usepackage{mathtools, nccmath}
\usepackage{dblfloatfix}
\usepackage{amsmath}
\usepackage{dsfont}
\usepackage{wrapfig}
\usepackage{algorithm}
\usepackage[noindentafter]{titlesec}
\usepackage{amsthm}
\usepackage{float}
\usepackage{booktabs}
\usepackage{hyperref}
\usepackage{bbm}
\usepackage{mathtools}
\usepackage{bbm}
\usepackage{algpseudocode}
\usepackage{mathtools}
\usepackage{amsmath,amssymb,amsfonts}
\usepackage{enumitem}
\usepackage{pifont}
\usepackage{xcolor}
\usepackage{stackengine}

\newcommand{\blue}[1]{\textcolor{black}{#1}}

\usepackage{bbding}
\usepackage{pifont}
\usepackage{wasysym}
\usepackage{amssymb}
\usepackage{amsthm}

\theoremstyle{definition}
\newtheorem{definition}{Definition}[section]

\begin{document}
\urlstyle{tt}

\title{GOSH: Task Scheduling Using Deep Surrogate Models in Fog Computing Environments}

\author{
        Shreshth~Tuli,
        Giuliano~Casale
    and~Nicholas~R.~Jennings
\IEEEcompsocitemizethanks{
\IEEEcompsocthanksitem S. Tuli, G. Casale and N. R. Jennings are with the Department
of Computing, Imperial College London, United Kingdom.\protect
\IEEEcompsocthanksitem N. R. Jennings is also with Loughborough University, United Kingdom.\protect\\
E-mails: \{s.tuli20, g.casale\}@imperial.ac.uk, n.r.jennings@lboro.ac.uk.\protect
}
\thanks{Manuscript received ---; revised ---.}}


\markboth{IEEE Transactions on Parallel and Distributed Systems}%
{Tuli \MakeLowercase{\textit{et al.}}: --- }

\IEEEtitleabstractindextext{%
\begin{abstract}
Recently, intelligent scheduling approaches using surrogate models have been proposed to efficiently allocate volatile tasks in heterogeneous fog environments. Advances like deterministic surrogate models, deep neural networks (DNN) and gradient-based optimization allow low energy consumption and response times to be reached. However, deterministic surrogate models, which estimate objective values for optimization, do not consider the uncertainties in the distribution of the Quality of Service (QoS) objective function that can lead to high Service Level Agreement (SLA) violation rates. Moreover, the brittle nature of DNN training and the limited exploration with low agility in gradient-based optimization prevent such models from reaching minimal energy or response times. To overcome these difficulties, we present a novel scheduler that we call GOSH for Gradient Based Optimization using Second Order derivatives and Heteroscedastic Deep Surrogate Models. GOSH uses a second-order gradient based optimization approach to obtain better QoS and reduce the number of iterations to converge to a scheduling decision, subsequently lowering the scheduling time. Instead of a vanilla DNN, GOSH uses a Natural Parameter Network (NPN) to approximate objective scores. Further, a Lower Confidence Bound (LCB) optimization approach allows GOSH to find an optimal trade-off between greedy minimization of the mean latency and uncertainty reduction by employing error-based exploration. Thus, GOSH and its co-simulation based extension GOSH*, can adapt quickly and reach better objective scores than baseline methods. We show that GOSH* reaches better objective scores than GOSH, but it is suitable only for high resource availability settings, whereas GOSH is apt for limited resource settings. Real system experiments for both GOSH and GOSH* show significant improvements against the state-of-the-art in terms of energy consumption, response time and SLA violations by up to 18, 27 and 82 percent, respectively.
\end{abstract}

}

\maketitle

\IEEEdisplaynontitleabstractindextext

\IEEEpeerreviewmaketitle


\IEEEraisesectionheading{\section{Introduction}\label{sec:introduction}}

\IEEEPARstart{M}{ainstream} parallel and distributed computing (PDC) technologies like Edge and Cloud Computing have recently been combined together into a hybrid paradigm commonly termed Fog Computing. Fog computing architectures are currently one of the most successful technologies to seamlessly harness both remote computational cloud resources, as well as compute capabilities close to the edge of the network~\cite{yousefpour2019all}. Such environments use containers for lightweight, flexible and fine-grained resource sharing among the myriad of fog devices~\cite{tuli2019fogbus}. However, scheduling tasks in heterogeneous fog environments with volatile workloads is challenging~\cite{tuli2021cosco}. Moreover, modern day users demand ultra-low response times and energy consumption, thanks to the shift to AI/ML/DL based workloads~\cite{liang2020ai, tuli2021generative}. Finally, resource heterogeneity further complicates task placement in such fog environments due to non-uniform task affinity for optimal cost-performance trade-off~\cite{chang2019internet}.

\textbf{Existing Solutions.} To deal with the challenges of scheduling volatile tasks in heterogeneous fog environments, recent work uses dynamic scheduling strategies which adapt to changing infrastructure conditions in real-time~\cite{matrouk2021scheduling}. Many schedulers use techniques like reinforcement learning and update their estimate of the ``expected reward'' for dynamic optimization of the Quality of Service (QoS)~\cite{basu2019learn, tuli2020dynamic, ghosal2020deep}. But, due to brittle modeling assumptions, high scheduling times and slow learning, they are not always suitable for highly non-stationary applications~\cite{tuli2021cosco}. To overcome these limitations, other approaches use max-weight based strategies with confidence-based exploration and virtual-queues for scheduling in constrained environments~\cite{liu2020pond}. Still, recent work using gradient based optimization and a deterministic deep surrogate model\footnote{A deep surrogate model is defined as a method used when an outcome of interest, like QoS score, cannot be easily directly measured. Examples include neural approximators and stochastic processes~\cite{tuli2021cosco}. In this work, we only refer to DNN-based surrogate models; hence, the term \textit{deep surrogate models}.} has been shown to outperform such approaches~\cite{tuli2021cosco}. Advances like deterministic surrogate models, deep neural networks (DNN) and gradient-based optimization allow low energy consumption and response times to be reached. This is due to their ability to quickly adapt to diverse scenarios in real-time.  However, due to the lack of state-space exploration or uncertainty modeling, such methods can perform poorly in settings unseen during model training~\cite{panda2015uncertainty}. Moreover, QoS optimization surfaces are known to be highly non-convex, and ignoring their curvature information can lead such methods to get stuck in local optima, saddle points or plateaus~\cite{tychogiorgos2013non}.

\begin{table*}[]
    \centering
    \caption{Comparison of related works with different parameters (\checkmark means that the corresponding feature is present).}
    \resizebox{\textwidth}{!}{
    \begin{tabular}{@{}lccccccccccc@{}}
    \toprule 
    \multirow{2}{*}{Work} & \multicolumn{1}{c}{Edge} & Heterogeneous & Coupled & Stochastic & Adaptive & Second-Order & Stochastic & \multirow{2}{*}{Method} & \multicolumn{3}{c}{Optimization Parameters}\tabularnewline
    \cline{10-12}
     & Cloud & Environment & Simulation & Workload & QoS & Optimization & Modelling &  & Energy & Response Time & SLA Violations\tabularnewline
    \midrule 
    \cite{zhang2018double, basu2019learn, gazori2019saving} &  & \checkmark &  &  &  &  &  & Deep RL & \checkmark &  & \tabularnewline
     
    \cite{tuli2020dynamic, ghosal2020deep} & \checkmark & \checkmark &  & \checkmark & \checkmark &  &  & Policy Gradient & \checkmark & \checkmark & \checkmark\tabularnewline
     
    \cite{liu2020pond, bae2019beyond, krishnasamy2018augmenting} &  & \checkmark &  & \checkmark & \checkmark &  &  & MaxWeight &  & \checkmark & \tabularnewline

    \cite{panda2015uncertainty, gruian2003uncertainty} &  &  &  & \checkmark &  &  & \checkmark & Heuristics & \checkmark & \checkmark & \tabularnewline

    \cite{zinnen2011deadline, jamshidi2016uncertainty, bui2017energy} &  &  &  &  & \checkmark &  & \checkmark & HGP & \checkmark & \checkmark & \tabularnewline
     
    \cite{tuli2021cosco} & \checkmark & \checkmark & \checkmark & \checkmark & \checkmark &  &  & GOBI/GOBI{*} & \checkmark & \checkmark & \checkmark\tabularnewline
     
    \textbf{This work} & \checkmark & \checkmark & \checkmark & \checkmark & \checkmark & \checkmark & \checkmark & GOSH/GOSH{*} & \checkmark & \checkmark & \checkmark\tabularnewline
    \bottomrule 
    \end{tabular}
    }
    \label{tab:related_works}
\end{table*}

\textbf{New Insights.} To overcome the challenges of unseen settings, it is important to consider uncertainty in the model predictions for a guided exploration of the state-space. In general, uncertainty may arise from several factors, including model approximations, measurement imprecision and fluctuations of parameters over time~\cite{tchernykh2015towards}. Specifically, having a deterministic model does not always give a complete picture of the distribution, due to the unknown nature or inherent stochasticity of some parameters like network latency, temperature and hardware faults~\cite{tuli2021pregan}. Thus, in this work, we use heteroscedastic surrogate models, specifically the Natural Parameter Network (NPN) that uses twice the parameters of a traditional DNN to output both mean and variance estimates. Further, it is crucial to model the curvature of the QoS hyper-surface to converge quickly to an optima~\cite[\S~6.1]{kochenderfer2019algorithms}. This entails using higher-order derivatives of the surrogate output while running decision optimization. Further, when modeling uncertainty, we may leverage this to trade-off between the exploration and exploitation to converge to the optimal scheduling decision~\cite{aima}. However, even after feeding more data, the performance of deep neural networks tends to saturate and we use recently proposed co-simulation based bootstrapping schemes to circumvent this~\cite{tuli2021cosco}. A co-simulator is a discrete-event simulator that can provide schedulers with future QoS score estimates for more robust scheduling~\cite{tuli2021cosco}.


\textbf{Contributions.} Drawing these insights together, we present \textbf{GOSH:} \textbf{G}radient Based \textbf{O}ptimization using \textbf{S}econd Order derivatives and \textbf{H}eteroscedastic Surrogate Model. We perform extensive empirical experiments on a real-life fog computing testbed to compare and analyze GOSH against state-of-the-art methods. Our experiments show that GOSH performs \textit{best} in terms of QoS metrics, reducing the energy consumption, response time and Service Level Agreement (SLA) violations by up to 18, 27 and 82 percent, respectively. It achieves this by being one of the \textit{first} methods to mitigate volatility by fully taking into account the uncertainty in DNN model predictions. Specifically, the technical contributions of this work include the use of a heteroscedastic neural network based surrogate model of the QoS objective score, uncertainty-based exploration and an adapted version of higher-order optimization. An extension of GOSH, namely GOSH*, further improves model predictions by running a discrete-event co-simulator to generate QoS estimates, but at the cost of increasing the scheduling time. 

The rest of the paper is organized as follows. Section~\ref{sec:related_work} overviews related work.  Section~\ref{sec:heteroscedastic} provides the system model assumptions, the working of the heteroscedastic models, uncertainty based exploration and higher-order optimization methods. Section~\ref{sec:gosh} presents the GOSH and GOSH* schedulers. A performance evaluation of the proposed methods on physical and simulated fog environments is shown in Section~\ref{sec:perf_eval}. Finally, Section~\ref{sec:conclusions} concludes and presents future directions.


\section{Related Work}
\label{sec:related_work}

We list in Table~\ref{tab:related_works} the prior work, dividing such methods into four classes: Reinforcement learning (rows 1 and 2 of Table~\ref{tab:related_works}), MaxWeight (row 3), stochastic modelling (rows 4 and 5) and gradient-based methods (row 6). 

\textbf{Reinforcement Learning models:} Many recent scheduling approaches use various forms of reinforcement learning to adapt in hybrid edge-cloud environments. Predominantly, these approaches use policy-gradient methods, which use a neural network to model a policy and output a decision for each state (assuming Markovian dynamics). In a dynamic scheduling setup, at each scheduling interval, a reward signal that quantifies the QoS of that interval is used to update the parameters of the model. Examples include asynchronous-actor-critic (A3C) policy gradient approach~\cite{tuli2020dynamic}. Such methods are shown to be robust and able to adapt in diverse scenarios~\cite{basu2019learn, tuli2020dynamic, ghosal2020deep}. However, such methods can be slow to adapt and fail to model volatile scenarios/workloads to be able to efficiently schedule in complex environments~\cite{tuli2021cosco}. Furthermore, they are known to perform poorly in heterogeneous setups or when the QoS is adaptive~\cite{nandi2001artificial}. 

\textbf{MaxWeight based schedulers:} Many prior scheduling approaches use the popular MaxWeight based techniques with robust theoretical guarantees. Such strategies aim to maximize the "weight" of job schedules where this weight is calculated using expected QoS metrics~\cite{liu2020pond, krishnasamy2018augmenting, bae2019beyond}. To compare with our method, we use the POND approach as a baseline due to its high performance compared to other methods in this class~\cite{liu2020pond}. POND uses a pessimistic-optimistic Upper Confidence Bound (UCB) based optimization in maximization setup. UCB optimization of the weight of virtual queues for constrained online dispatch minimizes the violations and maximizes the expected average reward. However, such techniques are unable to leverage the edge-cloud diversity in energy consumption and latency to improve SLA~\cite{panda2015uncertainty}.

\textbf{Stochastic Modelling:} Prior work also uses stochastic surrogate models such as Heteroscedastic Gaussian Processes to optimize metrics like QoS by appropriate task scheduling~\cite{zinnen2011deadline, jamshidi2016uncertainty, bui2017energy, panda2015uncertainty}. Other works use mean and variance estimates based on historical data to perform safe optimization~\cite{panda2015uncertainty, gruian2003uncertainty}. These methods typically utilize the uncertainty information for robust or safe optimization~\cite{jawad2018robust} or use error-based exploration~\cite{jamshidi2016uncertainty}. Typically, due to poor modeling accuracy of Gaussian Processes, they are unable to perform well in complex environments like heterogeneous fog environments and hence more sophisticated models like NPNs are used in this work~\cite{da2018resource}.


\textbf{Gradient-based Optimization:} A recently proposed class of methods (GOBI/GOBI*) use a deep neural network based surrogate model to approximate the QoS parameters like energy and response time~\cite{tuli2021cosco}. An extension of the GOBI scheduler for long-term QoS estimation is the MCDS approach~\cite{tuli2021mcds} that utilizes Monte-Carlo simulations to estimate the reward at a future system state. Their predictions are aided by a co-simulated digital-twin of the target hosts which receive the tasks. These algorithms use gradient-based optimization approaches like AdamW~\cite{adamw} with cosine annealing and restarts~\cite{pan2015annealed,loshchilov2016sgdr} to converge to an appropriate scheduling decision. Leveraging gradient-based directed search on neural approximator with coupled-simulation, GOBI* is able to adapt to non-stationary workloads. The neural approximator used in these methods is a Fully Connected Network (FCN) with input including the host and task utilization metrics of Instructions per second (IPS), RAM, Disk and Bandwidth with the scheduling decisions for each active task. GOBI*'s neural approximator also includes the objective score after the single-step execution of GOBI's scheduling decision on a simulated platform. The output of these neural approximators is an objective score that is minimized.

Although GOBI* provides distinct advantages compared to the state-of-the-art, it also faces drawbacks. First, the model require a neural approximator to be pre-trained on several hours of workload traces generated using a random allocation based scheduler. This might be infeasible for application scenarios where swift deployment is required. Second, the deployment environment might be very different from the setup used to generate the training data for the model. Even though the model is able to adapt to completely different setups, experiments show that GOBI* lacks the agility required for QoS efficient scheduling (Section~\ref{sec:agility}).  Finally, even with techniques like momentum, our experiments show that the gradient-based optimization can take a large number of iterations to converge due to many saddle points in the QoS hyper-surface. Considering these shortcomings, there is a need for a model that is able to leverage state-space exploration to quickly adapt to setups different from the one used for training.  To this end, we present the GOSH approach as part of this work.

\section{Modelling and Exploration}
\label{sec:heteroscedastic}

To model the variance of the QoS objective score in our system, we characterize its uncertainties into two classes defined below.
\theoremstyle{definition}
\begin{definition}[Aleatoric Uncertainty]
The aleatory uncertainty of a collection of datapoints refers to an inherent variation in the physical system to be modeled, also known as irreducible uncertainty~\cite{shaker2020aleatoric}.
\end{definition}
\theoremstyle{definition}
\begin{definition}[Epistemic Uncertainty]
The epistemic uncertainty of a collection of datapoints refers to the deficiencies caused by a lack of knowledge or information, also known as reducible uncertainty~\cite{shaker2020aleatoric}.
\end{definition}

Both uncertainties are used in GOSH as discussed in Section~\ref{sec:heteroscedastic}. We now describe the system, workload and the QoS objective models that are used in this work.

\begin{figure}[]
    \centering
    \includegraphics[width=0.95\linewidth]{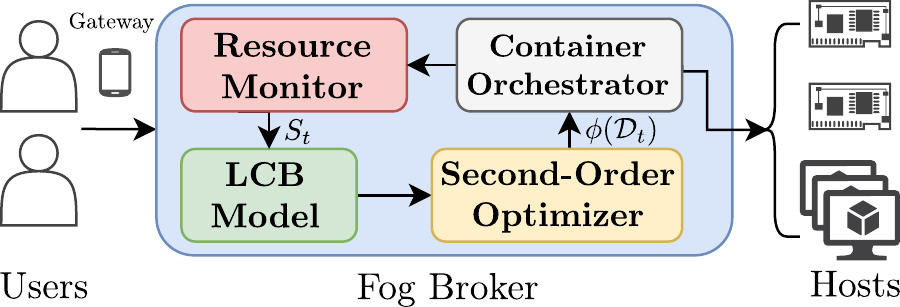}
    \caption{System Model.}
    \label{fig:system}
\end{figure}

\subsection{System Model}

We consider a distributed heterogeneous fog computing environment where tasks are generated from sensors and are relayed to the fog brokers via gateway nodes. Tasks manifest as containers that execute on fog worker nodes, also referred to as "hosts". Hosts at the edge are resource-constrained but offer low communication latency, whereas the ones in the cloud are equipped with heavy resources with high latency. The management and scheduling of all tasks is carried out by the fog broker, and the work described in this paper is concerned with the improvement of the scheduler in the broker.

We consider a bounded timeline, divided into equal sized scheduling intervals, with the $t$-th interval denoted as $I_t$. Moreover, we consider new tasks being generated at the beginning of every interval. For all tasks in the system in $I_t$, the broker takes a decision $\mathcal{D}_t$ which includes allocations for new and migrations for existing tasks (with matrix form denoted as $\phi(\mathcal{D}_t)$).The matrix form of a scheduling decision consists of one-hot vectors for each task allocation, each of the size of the number of hosts. Tasks unable to be allocated are kept in a wait queue $W_t$. The objective score at $I_t$ is denoted as $\mathcal{O}_t$. This objective score quantifies the QoS of the system.  The state of the system including the utilization metrics of hosts and tasks is denoted as $S_t$. Moreover, we train and use a Long-Short-Term-Memory (LSTM) network to predict the next state $S_{t+1}$ from the previous history of states $\{S_0, \ldots, S_t\}$~\cite{hochreiter1997long}. This is due to the ability of LSTMs to accurately model sequential data like utilization traces~\cite{tuli2021cosco}.

As shown in Figure~\ref{fig:system}, users generate tasks that are relayed via gateway devices to a fog broker. The broker uses the utilization state of the system ($S_t$) to generate an LCB estimate using a surrogate model as discussed in Section~\ref{sec:heteroscedastic}. This estimate is then optimized using a second-order gradient optimizer, and the final scheduling decision is executed by the container-orchestrator, discussed in Section~\ref{sec:second_order}. We now describe the neural model used by GOSH to model the QoS objective score $\mathcal{O}_t$.

As discussed in the introduction, it is crucial to model the complete distribution to characterize the uncertainty in metrics like response time to reduce SLA violations. To achieve this, as discussed in the following subsection, we use deep neural models of NPN due to their superior modeling performance to other methods like Gaussian Processes, BNNs or FCNs. Our optimization paradigm is motivated by energy-based models (EBMs) that use a deep neural network to model an energy function that is minimized using gradient-based optimization~\cite{lecun2007energy}. Here, the energy function takes as inputs the observed (like host and workload characteristics) and the decision variables (like a schedule for tasks). We now describe how we model this using NPNs, LCB and second-order optimization methods.

\subsection{NPN as Heteroscedastic Surrogate Model}
\label{sec:npn}

\begin{figure}
    \centering 
    \includegraphics[width=0.85\linewidth]{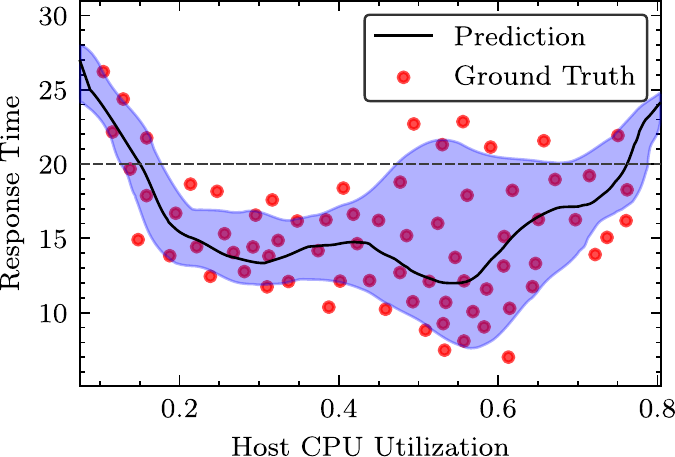}
    \caption{Heteroscedastic Surrogate for Response Time.}
    \label{fig:motivation}
\end{figure}

\textbf{Motivation.} Predicting just the mean response time using a deterministic surrogate model and ignoring the uncertainty in the distribution is insufficient for an accurate forecast of  Service Level Agreement (SLA) violations~\cite{panda2015uncertainty}.  Moreover, our tests show that in many cases (more than 70\%), taking a scheduling decision according to the minimum mean latency (but high uncertainty) gives a higher SLA violation rate than a decision with higher mean latency (but lower uncertainty). Consider the real-world experiments we have carried out in Figure~\ref{fig:motivation} where response time (seconds) is plotted against the CPU utilization of a host in a fog environment. Here, based on a dataset (red points), we train a heteroscedastic Gaussian process surrogate model (mean in black and confidence interval shaded). Considering the mean prediction model, there are two minima of the curve, 0.3 (local minimum) and 0.55 (global minimum) CPU utilization of a host. However, the probability of exceeding the SLA deadline (20 seconds) is higher for 0.55 than 0.3. Thus, both mean and statistical uncertainty need to be modeled to minimize the best objective value that can be guaranteed with some given probability viz the \textit{value at risk} (VaR)~\cite[\S~17.3]{kochenderfer2019algorithms}. 

\textbf{Natural Parameter Networks.} To model uncertainty, recent works have used models like Gaussian Processes (GP) or Bayesian Neural Networks (BNN)~\cite{zinnen2011deadline, jamshidi2016uncertainty}. However, prior work has shown Natural Parameter Networks (NPNs) to be more accurate in modeling heteroscedastic uncertainty, also corroborated by our experiments (Section~\ref{sec:sensitivity}). NPNs form a class of probabilistic-neural networks that provide a lightweight Bayesian treatment of traditional neural models~\cite{wang2016natural}. To do this, NPNs express neural network parameters as exponential-family distributions, giving a stochastic output of each layer. Such networks are trained using an adapted version of the backpropagation approach~\cite{wang2016natural}. Compared to standard BNNs, this allows NPNs to avoid executing time consuming sampling. Typically for an exponential distribution, the output can be characterized with a set of natural parameters (for instance, mean and variance in a normal distribution). Exploiting this, say for mapping a normal distribution, NPNs use twice the parameters of a traditional neural network to output both mean and variance estimates. In contrast to a traditional NN, an input goes through multiple layers of stochastic transformation, producing an output distribution that is matched with a target distribution. A major advantage of NPNs is that they model aleatoric uncertainty more accurately than the more popular Bayesian Neural Networks (BNNs)~\cite{wang2016natural}. However, inference using an NPN scales linearly with the dimension of the output~\cite{wang2016natural}.  Considering a uni-variate objective function, NPNs become a natural choice for stochastic modelling for our problem when compared to BNNs. 

\textbf{Heteroscedastic Modelling.} An NPN model cannot be used directly in our setting. This is because, unlike standard NPNs, we do not have the uncertainty information for each input state and scheduling decision. Instead, we have a single datapoint corresponding to each input of the network. To resolve this, we consider the underlying datapoint that corresponds to a probabilistic distribution instead of being static. Specifically, we consider a NPN neural approximator $f(x_t;\theta)$ of the objective score $\mathcal{O}$, with $\theta$ as the network parameters and $x_t$ as the state of the fog environment and scheduling decision $[S_t, \phi(\mathcal{D}_t)]$. The output of the NPN is a pair $(\mu_t, \sigma_t) = f(x_t,\theta)$. The parameters $\theta$ are learnt using a dataset $\Lambda = \{[S_t, \phi(\mathcal{D}_t)], \mathcal{O}_t\}_{t=0}^T$, and the aleatoric loss $\mathcal{L}$ defined as:
\begin{equation}
\label{eq:aleatoric_loss}
\mathcal{L}(f, \Lambda) = \frac{1}{T} \sum_{(x_t,y_t) \in \Lambda} \frac{(\mu_t - y_t)^2}{2 \sigma_t^2} + \frac{1}{2}\ln{\sigma_t^2},
\end{equation}
where $(\mu_t, \sigma_t) = f(x_t;\theta)$. This is the same as Kullback–Leibler divergence loss between $\mathcal{N}(\mu, \sigma)$ and $\mathcal{N}(y_t, 1)$~\cite{wang2016natural} of the ground-truth datapoint. Using this loss function, we train our NPN model to predict an estimate $\mu$ and aleatoric uncertainty $\sigma$. 

Now, we calculate the \textit{value at risk} using the prediction of the NPN model. The value at risk (VaR) is defined as the best objective value that can be achieved with $\alpha$ probability, which is equivalent to the $\alpha$ quantile of the distribution. Considering a Gaussian distribution and 95\% confidence bound as per~\cite{wang2016natural, kochenderfer2019algorithms, rezaei2020mean}, the parameterized VaR for interval $I_t$ is given by 
\begin{equation}
    \label{eq:vart}
    VaR_t = \mu + 1.65\sigma.
\end{equation}
Optimizing this metric in GOSH allows it to minimize the mean latency ($\mu$) as well as reduce uncertainty ($\sigma$) in response time, leading to a lower probability of overshooting deadline and hence lower SLA violation rates.





\subsection{LCB Exploration}

\textbf{Motivation.} Prior work shows that exploration of the state space based on the uncertainty estimates of heteroscedastic models can provide a performance boost in dynamic scenarios~\cite{jamshidi2016uncertainty}.  In particular, methods like $\epsilon$-greedy, softmax and Lower-confidence bound (LCB) have been popular~\cite{jamshidi2016uncertainty, liu2020improving}. However, due to the adaptability of LCB exploration, we use LCB to trade off between greedy minimization of the mean latency and error reduction by employing uncertainty-based exploration~\cite[\S~17.3.4]{kochenderfer2019algorithms}. However, it is known that uncertainty modeling in such NPNs only includes the aleatoric uncertainty~\cite{sahlin2021we}. Thus, in this work, we use the LCB metric with epistemic uncertainty. 

\textbf{Teacher-Student Network.} To implement the LCB exploration strategy, we need estimates of the epistemic uncertainty for each input $[S_t, \phi(\mathcal{D}_t)]$. Prior work leverages models like BNNs or Monte-Carlo Dropout to achieve this by sampling the output multiple times and quantifying the variance in the output to determine the epistemic uncertainty~\cite{postels2019sampling}. However, the measure of epistemic uncertainty obtained from this method is not in closed form and hence cannot be differentiated. Having closed-form solutions is better for second-order optimization as finite difference methods tend to be slow and imprecise~\cite[\S~2.3.2]{kochenderfer2019algorithms}, having a negative effect on the scheduling time. 

\begin{figure}[]
    \centering
    \includegraphics[width=\linewidth]{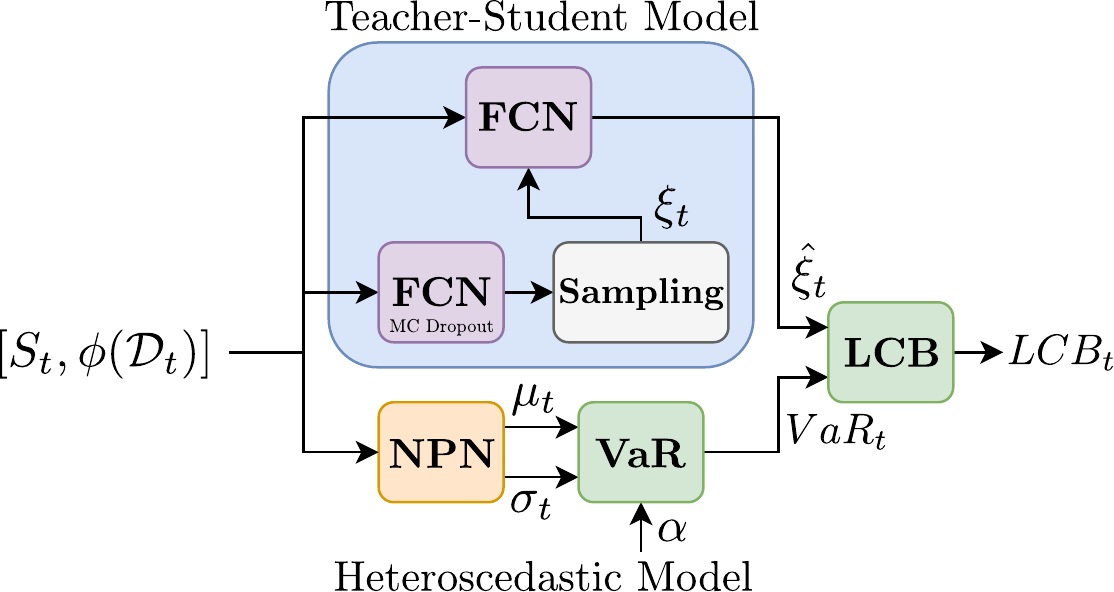}
    \caption{LCB metric calculation in GOSH.}
    \label{fig:lcb}
\end{figure}

To solve this problem, we employ a teacher-student learning model, similar to the one in~\cite{matiisen2019teacher}, with an overview given in Figure~\ref{fig:lcb}. Compared to other methods like approximated variance propagation or sampling based propagation~\cite{postels2019sampling}, a teacher-student model is able to adapt quickly to changing scenarios. \textbf{The teacher network}, denoted as $g(x_t; \theta')$, is a FCN model that takes $[S_t, \phi(\mathcal{D}_t)]$ as an input and outputs an estimate of $\mathcal{O}_t$. However, in this model, we use Monte-Carlo Dropout (MCD) for Bayesian inference at test time~\cite{gal2016dropout}. Unlike conventional dropout, MCD enables dropout at inference time as well. This allows us to run inference multiple times (say $M$) and obtain a stochastic output. We can then calculate the standard-deviation of the resultant output samples. This deviation quantifies the epistemic uncertainty (denoted as $\xi$) of the model as described in~\cite{kendall2017uncertainties}. In contrast, \textbf{the student network}, denoted as $h(x_t; \theta'')$, is another identical FCN model without the MC Dropout. It takes $[S_t, \phi(\mathcal{D}_t)]$ as an input and directly outputs an estimate of the epistemic uncertainty $\xi$ (denoted as $\hat{\xi}$). Unlike the teacher, the output epistemic uncertainty of the student is differentiable, allowing quick and precise second-order optimization. Both teacher ($g$) and student ($h$) networks are trained using the Mean Square Error (MSE) loss such that for a dataset $\Lambda$ as described previously, 
\begin{gather}
\label{eq:mse_mcd}
    \mathcal{L}(g, \Lambda) = \frac{1}{T} \sum_{(x_t,y_t) \in \Lambda} (g(x_t; \theta') - y_t)^2\\
\label{eq:mse_epistemic}
    \mathcal{L}(h, \Lambda) = \frac{1}{T} \sum_{(x_t,y_t) \in \Lambda} (h(x_t; \theta'') - \xi_t )^2.
\end{gather}
Assuming we use affine activations, the student-network is a differentiable closed-form function; its output $\hat{\xi}$ is differentiable with respect to the input.

Now that we have the VaR from the NPN model and an estimate of the epistemic uncertainty, we use the LCB optimization where the final objective function to minimize becomes 
\begin{equation}
\label{eq:lcb}
    LCB_t = \underbrace{VaR_t}_{\text{exploitation}} - \underbrace{k \cdot \hat{\xi}_t}_{\text{exploration}},
\end{equation}
where $k$ is the exploration factor. 

Minimizing the first term aims to reduce the chance of an SLA violation (considering response time as part of $\mathcal{O}_t$) and can be termed the \textit{exploitation} part of the algorithm. Minimizing the second term, \textit{i.e.}, maximizing the epistemic uncertainty helps the model \textit{explore} in cases with limited data or knowledge. This interactive dynamic between the exploration and exploitation allows the GOSH scheduler to quickly adapt while maintaining low SLA violation rates. 

\subsection{Second-Order Optimization}
\label{sec:second_order}

We now describe how we optimize the scheduling decision. To reduce the number of iterations, avoid saddle points and shorten the scheduling time, we employ a second-order stochastic optimization approach which dynamically incorporates the curvature of the surrogate function~\cite{yao2021adahessian}. Exploiting the improved convergence properties of the second order algorithms over first order ones, our approach gives \textit{low scheduling times} and \textit{improved QoS objective scores}.

For a trained neural approximator $f$, at the start of each interval $I_t$, a typical second-order approach (Newton methods) optimizes the decision matrix $\phi(\mathcal{D})$ by the following rule ($\gamma$ is the learning rate)
\[\phi(\mathcal{D}_{n+1}) \gets \phi(\mathcal{D}_{n}) - \gamma \cdot [\nabla^2_{\phi(\mathcal{D}_{n})}f]^{-1} \cdot \nabla_{\phi(\mathcal{D}_n)}f,\]
where $f$ is a shorthand for $f([S_t, \phi(\mathcal{D}_n)]; \theta)$. This operation is iteratively performed until the absolute value of the gradient is more than the convergence threshold $\epsilon$.

Computing the second-order derivative, or Hessian in a multivariate case, allows us to make a quadratic approximation of the objective function and estimate an appropriate step size to reach a local minimum~\cite[\S~6]{kochenderfer2019algorithms}. For instance, it is known that if $f$ is quadratic and its Hessian is positive definite, one can reach its global minimum in a single step using the Newton method. Moreover, second-order methods provide resilience against ill-conditioned objective function landscapes and guarantee convergence to second-order critical points~\cite{yao2021adahessian}.

However, computing the inverse of the Hessian is known to be computationally expensive (cubic time-complexity with matrix size). In this endeavor, prior work proposed Hessian free approaches like secant or Quasi-Newton methods~\cite[\S~6.3]{kochenderfer2019algorithms}. Moreover, it has been shown that the exact computation of Hessian is not required and an approximate of the diagonal of the Hessian is sufficient for low-cost iterations~\cite{martens2015optimizing}. One such method for computing this Hessian diagonal for neural networks is the Hutchinson's method with a linear computational complexity~\cite{yao2018hessian}. However, even an approximate Hessian can give local curvature information, which is misleading. 
This is alleviated by adapting the Hutchinson's method and using the moving average of the approximate Hessian diagonal to speed up convergence. Thus, for Hutchinson's estimate of the Hessian diagonal of $f([S_t, \phi(\mathcal{D}_n)]; \theta)$ at $I_t$, denoted as $H_t(f)$, we compute the moving average $\hat{H}_t(f)$ as
\[\hat{H}_t(f) \gets \omega \cdot \hat{H}_t(f) +  (1 - \omega) \cdot H_t(f),\]
where $\omega$ is the momentum parameter. Hence, the update equation becomes
\begin{equation}
\label{eq:sgobi}
\phi(\mathcal{D}_{n+1}) \gets \phi(\mathcal{D}_{n}) - \gamma \cdot \hat{H}_t^{-1}(f) \cdot \nabla_{\phi(\mathcal{D}_{n})}f.
\end{equation}

Unlike prior efforts that utilize Hutchinson's method for lightweight second-order optimization like~\cite{yao2021adahessian, yao2020pyhessian}, we perform iteration based moving averages and not spatial averaging. This allows our model to be robust to sudden variations in reward metrics in volatile workload settings. The main overhead of this approach is the computation of the approximate of the Hessian diagonal using Hutchinson's method. Also, it is possible to improve the estimate of the Hessian diagonal by increasing the iteration count of Hutchinson's method. However, one-step iteration performs as well in practice, giving low per-iteration computational overhead compared to the GOBI approach~\cite{yao2021adahessian}. Comparisons of the scheduling overheads are shown later in Section~\ref{sec:overhead}. 




\section{The GOSH and GOSH* Schedulers}
\label{sec:gosh}

In the GOSH approach, we combine the LCB metric given by \eqref{eq:lcb} and second-order optimization in \eqref{eq:sgobi}. Thus, our update equation becomes
\begin{equation}
\label{eq:gosh}
\phi(\mathcal{D}_{n+1}) \gets \phi(\mathcal{D}_{n}) - \gamma \cdot \hat{H}_t^{-1}(LCB_t) \cdot \nabla_{\phi(\mathcal{D}_{n})}LCB_t,
\end{equation}
where $\hat{H}_t(LCB_t)$ denotes the moving average of the Hessian diagonal estimate of $LCB_t$. As we utilize the output of the student network for the epistemic uncertainty, we can generate a closed-form second-order Hessian estimate (see Figure~\ref{fig:lcb}).

To optimize the QoS parameters, we consider a standard objective function that focuses on energy consumption and response time as is done in prior work~\cite{tuli2021cosco}. For interval $I_t$,
\begin{equation}
    \label{eq:objective_function}
    \mathcal{O}_t = \alpha \cdot AEC_t + \beta \cdot ART_t,
\end{equation}
where $AEC_t$ and $ART_t$ are the average energy consumption of the fog environment in $I_t$ and average response time of tasks leaving at the end of $I_t$. Here, $\alpha$ and $\beta$ (such that $\alpha+\beta=1$) are hyper-parameters that can be set by users in line with the application requirements. Typically, in real-life settings, these values are around 0.5~\cite{tuli2021cosco, tuli2020dynamic}. For an energy-constrained setting, a higher $\alpha$ value is used, whereas for latency-critical tasks, a higher $\beta$ value is used.

We also extend GOSH to benefit from co-simulated results. GOSH* uses a simulator to execute the scheduling decision predicted by GOSH and the utilization metrics of the next state using an LSTM (Section~\ref{sec:heteroscedastic}). This allows GOSH* to make more informed decisions and hence reach better objective scores than GOSH, spending additional computational overhead in return for increased accuracy. We will show that GOSH is more suitable for environments with resource-constrained fog brokers, whereas GOSH* reaches a better objective score but requires heavier computational resources.

\subsection{Model Training}

As described previously in Section~\ref{sec:npn}, the models $f, g$ and $h$ are trained using a dataset $\Lambda = \{[S_t, \phi(\mathcal{D}_t)], \mathcal{O}_t\}_{t=0}^T$ as discussed in Section~\ref{sec:heteroscedastic}. The model $f$ is trained using the aleatoric loss~\eqref{eq:aleatoric_loss}, whereas $g, h$ are trained using MSE loss given in~\eqref{eq:mse_mcd} and~\eqref{eq:mse_epistemic}. We train using the AdamW optimizer~\cite{loshchilov2017decoupled}. Using these models, we obtain the LCB metric~\eqref{eq:lcb}. 

For GOSH*, we train models $f^*, g^*$ and $h^*$ which take input $[S_t, \bar{O}_t, \phi(\mathcal{D})]$, where $\bar{\mathcal{O}}_t$ is a simulated estimate of $\mathcal{O}_t$. We also train a next-state predictor using a LSTM network (Section~\ref{sec:heteroscedastic}). This model uses the state history to predict the next state, where the predicted state is denoted as
\[\bar{S}_t \gets LSTM(\{ S_{t'} \forall t'<t \}).\]
This LSTM model is also trained using the $\Lambda$ dataset. 

\begin{algorithm}[t]
    \begin{algorithmic}[1]
    \Require
    \Statex Pre-trained stochastic function approximator $f(x;\theta)$
    \Statex Pre-trained Teacher-Student models $g(x;\theta'), h(x; \theta'')$
    \Statex Dataset used for training $\Lambda$; Convergence threshold $\epsilon$
    \Statex Iteration limit $L$; Learning rate $\gamma$; LCB parameter $k$
    \Statex Value at Risk parameter $\alpha$
    \Procedure{Minimize}{$\mathcal{D}$, f, g, h, $z$}
        \State Initialize decision matrix $\phi(\mathcal{D})$; $i = 0$
        \State \textbf{do}
        \State \hspace{\algorithmicindent} $x \gets [z, \phi(\mathcal{D})]$ \Comment{Concatenation}
        \State \hspace{\algorithmicindent} $\mu, \sigma \gets f(x; \theta)$ \Comment{Stochastic Prediction}
        \State \hspace{\algorithmicindent} $\hat{\xi} \gets h(x; \theta'')$ \Comment{Uncertainty Prediction}
        \State \hspace{\algorithmicindent} $LCB \gets VaR(\mu, \sigma, \alpha) - k \cdot \hat{\xi}$ \Comment{LCB metric}
        \State \hspace{\algorithmicindent} $\phi(\mathcal{D}) \gets \phi(\mathcal{D}) - \gamma \cdot \hat{H}^{-1}(LCB) \cdot \nabla_{\phi(\mathcal{D})}LCB$
        \State \hspace{\algorithmicindent} $i \gets i + 1$
        \State \textbf{while} $|\hat{H}^{-1}(LCB) \cdot \nabla_{\phi(\mathcal{D})}LCB| > \epsilon$ and $i \leq L$
        \State Convert matrix $\phi(\mathcal{D})$ to scheduling decision $\mathcal{D}^*$
        \State \textbf{return} $\mathcal{D}^*$
    \EndProcedure
    \Procedure{Initialize}{scheduling interval $I_t$}
        \State \textbf{if} (t == 0)
        \State \hspace{\algorithmicindent} \textbf{return} random decision $\mathcal{D}$
        \State \textbf{return} $\mathcal{D}_{t-1}$ \Comment{Output for the previous interval}
    \EndProcedure
    \Procedure{Tune}{scheduling interval $I_t$}
        \State $x_t \gets [S_t, \phi(\mathcal{D}_t)]$
        \State $\xi_t \gets Sample(g(x_t ; \theta'))$
        \State $k, \xi_{MA} \gets Update(\xi, \psi)$
        \State $\Lambda_{t-1} \gets \{(x_{t-1}, \mathcal{O}_{t-1})\}$
        \State Fine-tune $\theta, \theta'$ with $\Lambda_{t-1}$ and loss as in \eqref{eq:aleatoric_loss}, \eqref{eq:mse_mcd}
        \State Fine-tune $\theta''$ with $\xi_t$ and loss as in \eqref{eq:mse_epistemic}
    \EndProcedure
    \Procedure{GOSH}{scheduling interval $I_t$}
        \State $\mathcal{D} \gets  \textsc{Initialize}(I_t)$
        \State Get $S_t$
        \State $\mathcal{D}_t \gets \textsc{Minimize}(\mathcal{D}, f, g, h, S_t)$
        \State $\textsc{Tune}(I_t)$
        \State \textbf{return} $\mathcal{D}_t$
    \EndProcedure
    \end{algorithmic}
\caption{The GOSH scheduler}
\label{alg:gosh}
\end{algorithm}

\subsection{Dynamic Exploration Weight}

As we show in Section~\ref{sec:sensitivity}, keeping a static value of the exploration weight $k$ is not ideal as the objective scores are very sensitive to its value. Hence, GOSH and GOSH* use a dynamic exploration weight $k$, which is updated based on the epistemic uncertainty value in each scheduling interval. Thus, we update the $k$ parameter such that it increases when the epistemic uncertainty increases, indicating more exploration required and decreases otherwise, while at the same time remaining fairly insensitive to the step-size parameter.

We use a feedback based weight control scheme similar to the one in~\cite{maroti2019rbed}. At run-time, we maintain an exponential moving average of the epistemic uncertainty metric $\xi$ (denoted as $\xi_MA$). Thus, at each interval $I_t$, $\xi_{MA} \gets \psi \cdot \xi_{MA} + (1-\psi) \cdot \xi_t$, where $\psi$ is the exploration momentum parameter. Now, the value of $k$ is updated as per the rule
\begin{equation}
\label{eq:dynamic_k}
k =
\begin{cases}
  (1+\delta) \cdot k, & \text{if}\ \xi > (1+\delta) \cdot \xi_{MA} \\
  (1-\delta) \cdot k, & \text{if}\ \xi < (1+\delta) \cdot \xi_{MA} \\
  k, & \text{otherwise},
\end{cases}
\end{equation}
where $\delta$ is the step-size.

\subsection{Scheduling}

\begin{algorithm}[t]
    \begin{algorithmic}[1]
    \Require
    \Statex Pre-trained stochastic function approximator $f^*(x;\theta)$
    \Statex Pre-trained Teacher-Student models $g^*(x;\theta'), h^*(x; \theta'')$
    \Statex Pre-trained LSTM model $LSTM(\{ S_{t'} \forall t'<t \})$
    \Procedure{$\text{GOSH}^*$}{scheduling interval $I_t$}
        \State $\mathcal{D} \gets  \textsc{Initialize}(I_t)$
        \State Get $S_t$
        \State $\bar{S}_t \gets LSTM(\{ S_{t'} \forall t'<t \})$
        \State $\bar{\mathcal{D}}_t \gets GOSH(I_t)$
        \State $\bar{\mathcal{O}}_t \gets \mathcal{S}(\bar{\mathcal{D}}(A_t),  \bar{S}_t)$ \Comment{Simulation}
        \State $\mathcal{D}_t \gets \textsc{Minimize}(\mathcal{D}, f^*, g^*, h^*, [S_t, \bar{\mathcal{O}}_t])$
        \State $\textsc{Tune}(I_t)$
        \State \textbf{return} $\mathcal{D}_t$
    \EndProcedure
    \end{algorithmic}
\caption{The GOSH* scheduler}
\label{alg:gosh2}
\end{algorithm}

An overview of the GOSH scheduler is given in Algorithm~\ref{alg:gosh}.
After training models $f, g$ and $h$ and obtaining LCB metric, we optimize $\phi(\mathcal{D}_t)$ as per the rule~\eqref{eq:gosh} until the step-size $\hat{H}_t^{-1}(LCB_t) \cdot \nabla_{\phi(\mathcal{D}_{n})}LCB_t$ is greater than the convergence threshold $\epsilon$ (line 10).  We also use Implicit Gradient Regularization in the $\phi(\mathcal{D}_t)$ optimization steps for numerical stability in high aleatoric uncertainty cases~\cite{barrett2020implicit}. This adds a loss term that imposes structural constraints to the gradient values making them more robust to noisy parameter perturbations~\cite{barrett2020implicit}.

For interval $I_0$, we initialize a random decision and for $I_t, t>0$ we initialize the decision to be the same as $\mathcal{D}_{t-1}$ (lines 16-17). We obtain the current state of the system which includes the IPS, RAM, Disk and Bandwidth utilizations of the hosts and tasks as $S_t$ (line 29). We utilize the $\textsc{Minimize}$ function, given in Algorithm~\ref{alg:gosh}, to find the decision at each interval $I_t$ (line 30). We then fine-tune the models $f, g$ using the latest datapoint $(x_{t-1}, \mathcal{O}_{t-1})$ (line 24). Finally, we sample and obtain the epistemic uncertainty from $g$ as $\xi_t$ and fine-tune $h$ using MSE loss (line 25). This continuous fine-tuning allows the models to adapt in dynamic scenarios.

An overview of the working GOSH* scheduler in given in Algorithm~\ref{alg:gosh2}. Here, we employ GOSH, next-state predictor (LSTM model) and a simulator (denoted as $\mathcal{S}$). Using the predicted next-state $\bar{S}_t$ and an action of interest $\bar{\mathcal{D}}_t = GOSH(I_t)$, we use the simulator to predict the objective score at the end of the current interval  $\bar{\mathcal{O}}_t \gets \mathcal{S}(\bar{\mathcal{D}}(A_t),  \bar{S}_t)$ (line 6).


\section{Performance Evaluation}
\label{sec:perf_eval}


\subsection{Evaluation Setup}

To show the performance of the GOSH and GOSH* models against baselines methods, we use two test-beds, physical and simulated platforms. The former for precise comparison and the latter for large-scale experiments. For our experiments, we keep $\alpha = \beta = 0.5$ in~\eqref{eq:objective_function} and $\delta = 0.1$ in~\eqref{eq:dynamic_k} since the same values are used in the baselines we compare with~\cite{tuli2020dynamic, tuli2021cosco, maroti2019rbed}. These values have been shown to represent a characteristic set of deployments closely resembling the ones in practical settings~\cite{tuli2020healthfog, tuli2021hunter}. We get similar trends for other values.

\textbf{Physical Setup:} Motivated from prior work~\cite{tuli2021cosco, tuli2020dynamic}, we use Microsoft Azure VMs to create a heterogeneous fog test-bed. The configuration consists of 10 VMs and is as follows:
\begin{itemize}
    \item 4 Azure B2s (dual core) VMs in London, UK.
    \item 2 Azure B4ms (quad core) VMs in London, UK.
    \item 2 Azure B4ms (quad core) VMs in Virginia, USA.
    \item 2 Azure B8ms (octa core) VMs in Virginia, USA.
\end{itemize}
The fog broker node was a system situated in {London, UK} and with configuration: Intel i7-10700K CPU, 64GB RAM and Windows 10 Pro OS. All neural network models are trained in the broker node.

\textbf{Simulated Setup:} We consider 50 host machines as a scaled up version of the physical setup. Here, each category has five times the instance count to test models in a larger-scale fog environment as considered in prior art~\cite{basu2019learn, ahmed2018docker}. 

We run all experiments for 100 scheduling intervals, with each interval being 300 seconds long, giving a total experiment time of 8 hours 20 minutes. We average over five runs and use diverse workload types to ensure statistical significance. \blue{We truncate the first 10\% of the scheduling intervals to ensure that the system reaches a stable state and minimize noise in the results}.

\subsection{Workloads}

We use popular and real-world workloads to test GOSH and GOSH* models. For our simulator, we use the \textit{Bitbrain}\footnote{The BitBrain dataset can be downloaded from: \url{http://gwa.ewi.tudelft.nl/datasets/gwa-t-12-bitbrains}} workload traces dataset~\cite{shen2015statisticalBitBrain}. These traces are generated by running computational analysis applications on 1750 VMs. The workloads consist of data intervals of 5 minutes, including the CPU, RAM, Disk and Bandwidth utilization of all tasks. We use both \textit{random} and \textit{sequential} computation traces, sampling them uniformly at random when generating workload containers.

For our physical setup, we use \textit{DeFog} benchmark applications~\cite{mcchesney2019defog}. We use the COSCO container orchestration platform~\cite{tuli2021cosco} to map these fog benchmark applications to containers run in our physical setup. These include the popular deep learning \blue{bag-of-task} applications: Yolo, PocketSphinx and Aeneas. At the beginning of each scheduling interval we create $Poisson(\lambda)$ tasks with $\lambda = 1.2$ tasks for physical and $\lambda = 5$ tasks for the simulated setup~\cite{tuli2020dynamic}.

\begin{figure*}
    \centering 
    \subfigure[Average Energy Consumption]{
    \includegraphics[width=.235\textwidth]{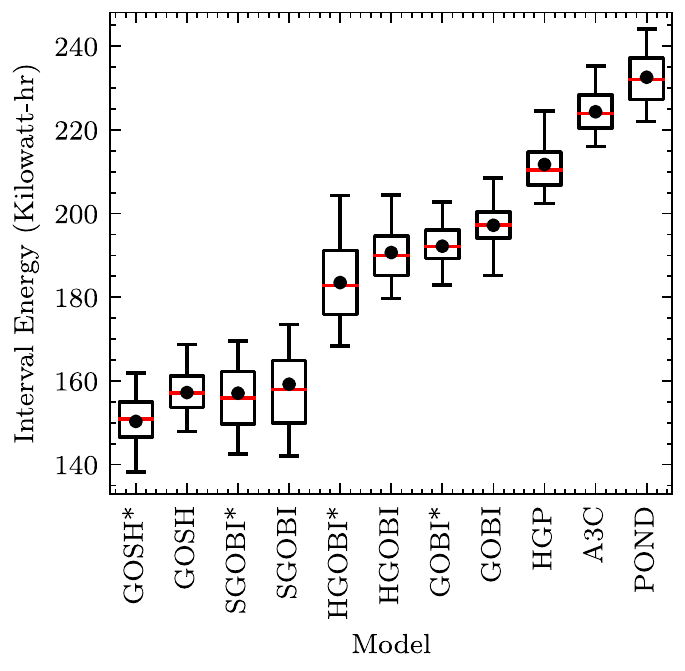}
    \label{fig:f_energy}
    }
    \subfigure[Average Response Time]{
    \includegraphics[width=.235\textwidth]{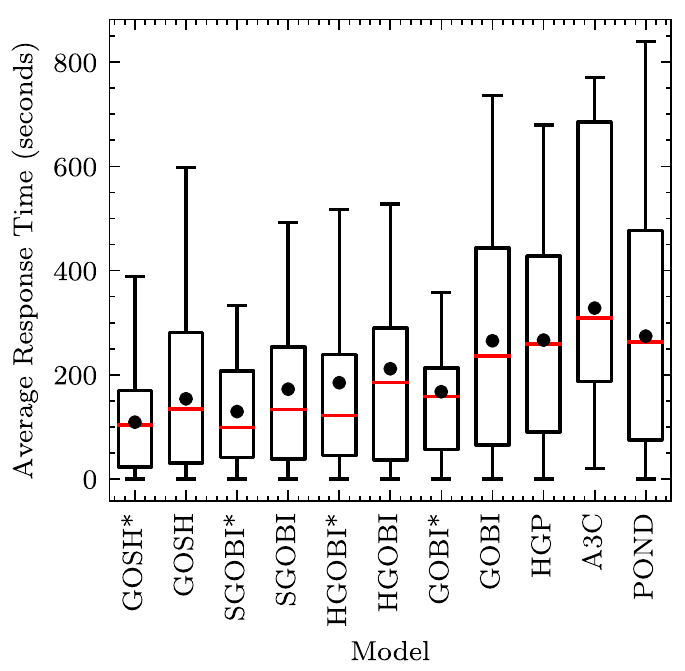}
    \label{fig:f_response}
    }
    \subfigure[Fairness]{
    \includegraphics[width=.235\textwidth]{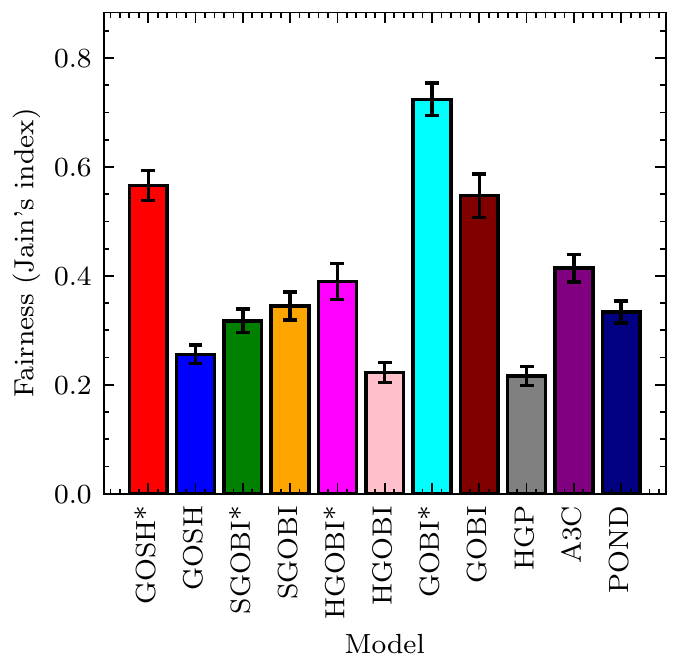}
    \label{fig:f_fairness}
    }
    \subfigure[Fraction of SLA Violations]{
    \includegraphics[width=.235\textwidth]{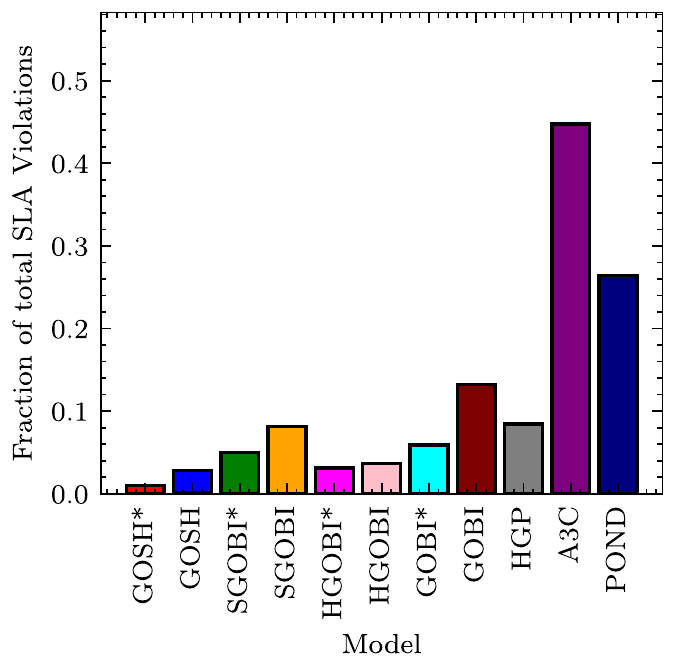}
    \label{fig:f_sla}
    }\\
    \subfigure[Average Migration Time]{
    \includegraphics[width=.235\textwidth]{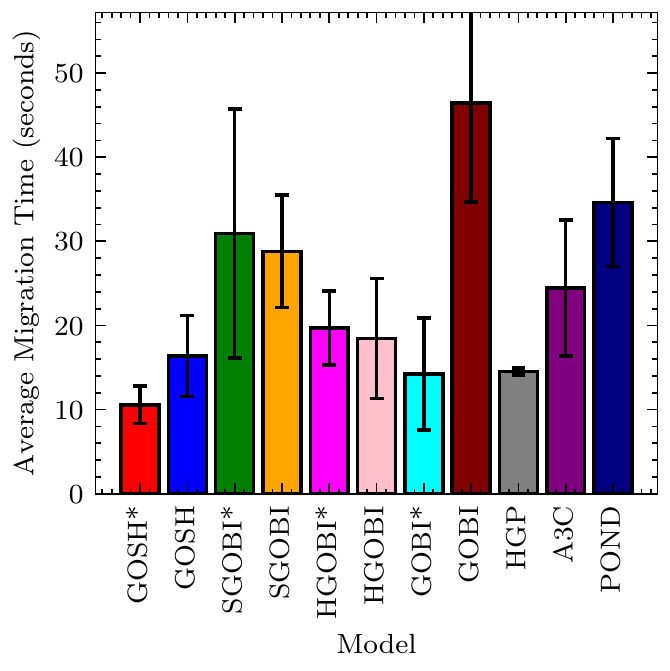}
    \label{fig:f_migration_time}
    }
    \subfigure[Average Scheduling Time]{
    \includegraphics[width=.235\textwidth]{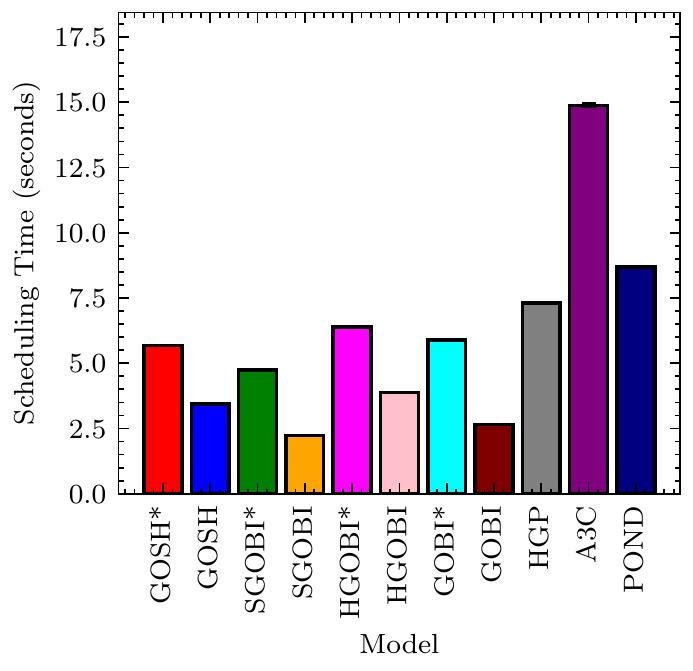}
    \label{fig:f_scheduling_time}
    }
    \subfigure[Average Response Time (per application)]{
    \includegraphics[width=.235\textwidth]{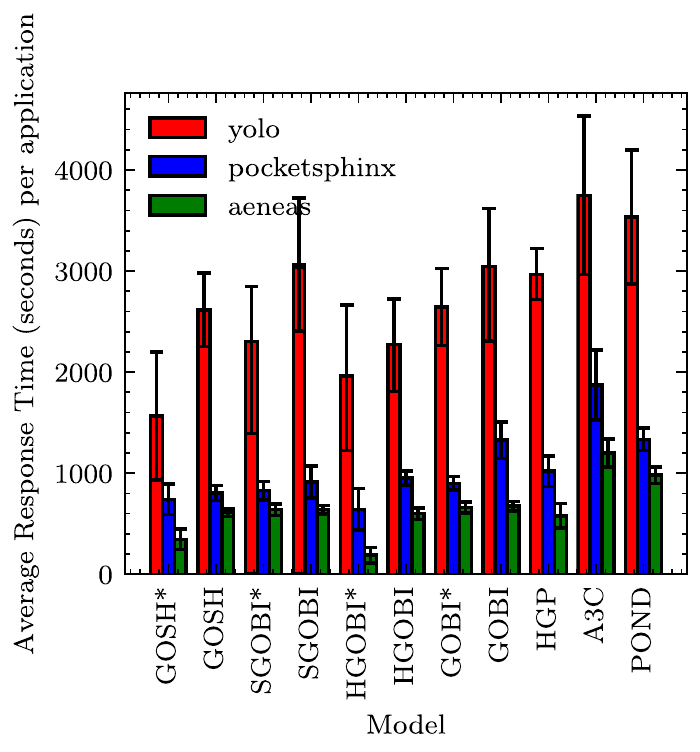}
    \label{fig:f_response_pa}
    }
    \subfigure[Average SLA violations (per application)]{
    \includegraphics[width=.235\textwidth]{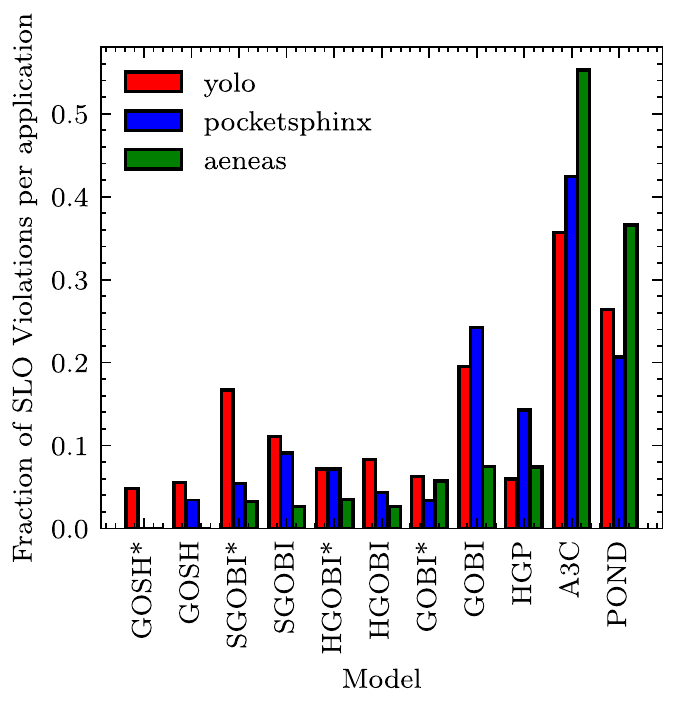}
    \label{fig:f_sla_pa}
    }\\
    \includegraphics[width=.85\textwidth]{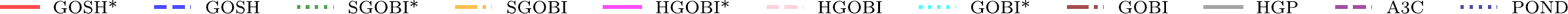}\\
    \subfigure[Average energy with intervals]{
    \includegraphics[width=.235\textwidth]{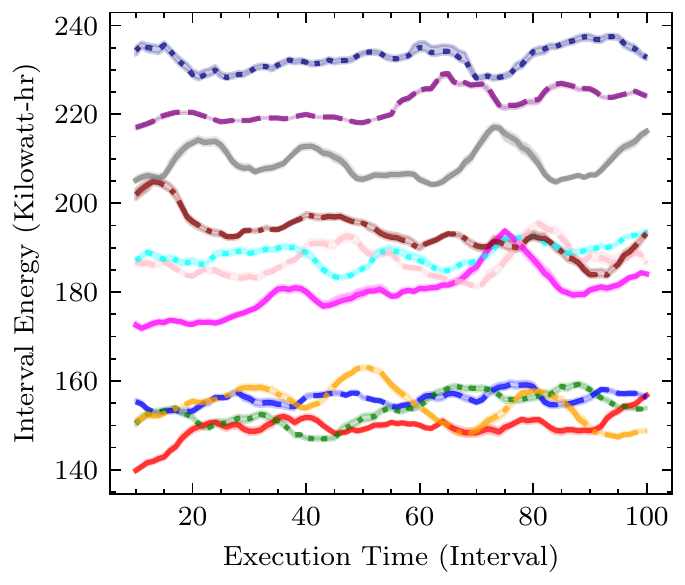}
    \label{fig:f_energy_series}
    }
    \subfigure[Scheduling time with intervals]{
    \includegraphics[width=.235\textwidth]{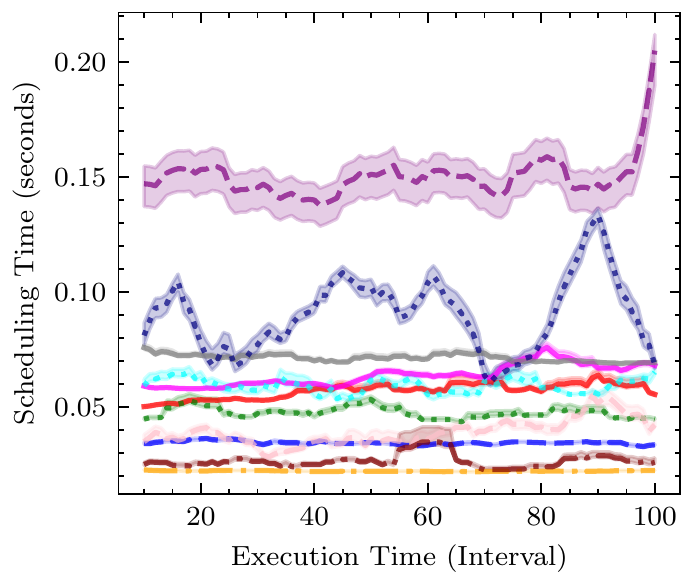}
    \label{fig:f_scheduling_time_series}
    }
    \subfigure[Average wait time with intervals]{
    \includegraphics[width=.235\textwidth]{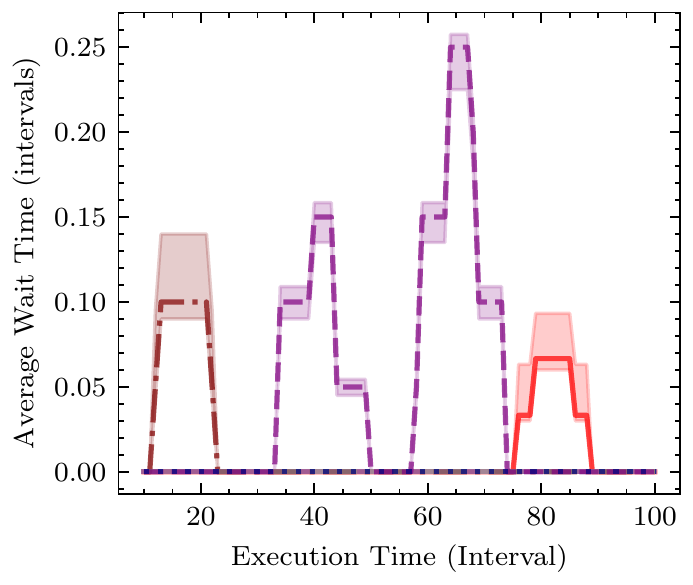}
    \label{fig:f_wait_time_series}
    }
    \subfigure[Average number of containers per host with intervals]{
    \includegraphics[width=.235\textwidth]{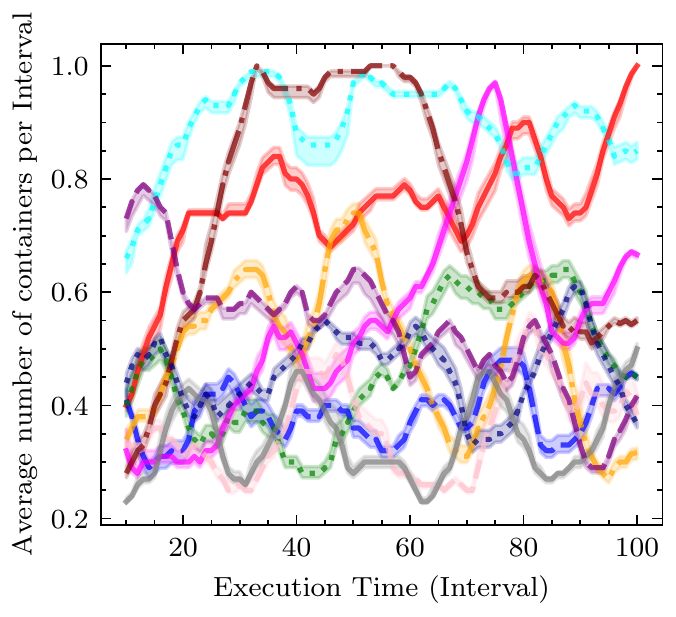}
    \label{fig:f_containers_series}
    }
    \caption{\blue{Comparison of GOSH and GOSH* against baselines and ablated models on physical setup with 10 hosts}}
    \label{fig:framework_results} \vspace{-1.2em}
\end{figure*}

\begin{table*}[t]
\centering
\caption{\blue{Comparison of GOSH and GOSH* against baselines and ablated models on simulator with 50 hosts. The best achieved value for each metric is shown in bold. Standard deviations are also indicated for metrics averaged over all completed tasks.} }
\begin{tabular}{@{}lccccccc@{}}
\toprule 
\textbf{Model} & \textbf{Energy} & \textbf{Response Time} & \textbf{SLA violation} & \textbf{Scheduling Time} & \textbf{Fairness} & \textbf{Migration Time} & \textbf{Wait Time}\tabularnewline
\midrule
POND & 326.21 & 2083.81$\pm$176.25 & 2.93E-01 & 18.94$\pm$0.01 & 0.56$\pm$0.01 & 1.10$\pm$0.07 & 114.28$\pm$0.32\tabularnewline
A3C & 270.28 & 2017.52$\pm$164.76 & 3.29E-01 & 15.77$\pm$0.03 & 0.51$\pm$0.02 & 0.72$\pm$0.06 & 206.34$\pm$0.26\tabularnewline
HGP & 364.20 & 2219.36$\pm$202.75 & 1.32E-01 & 21.56$\pm$0.00 & \textbf{0.60$\pm$0.01} & 0.21$\pm$0.01 & 60.64$\pm$0.12\tabularnewline
GOBI & 214.29 & 1767.29$\pm$126.00 & 1.51E-01 & 6.19$\pm$0.00 & 0.53$\pm$0.01 & 0.84$\pm$0.01 & 26.98$\pm$0.07\tabularnewline
GOBI{*} & 211.05 & 1487.06$\pm$123.98 & 7.02E-02 & 13.73$\pm$0.00 & 0.59$\pm$0.01 & 0.31$\pm$0.02 & \textbf{4.13$\pm$0.04}\tabularnewline
\midrule 
HGOBI & 209.22 & 1557.84$\pm$137.79 & 5.22E-02 & 11.22$\pm$0.01 & 0.53$\pm$0.01 & 0.38$\pm$0.02 & 24.90$\pm$0.11\tabularnewline
HGOBI{*} & 215.29 & 1295.32$\pm$157.90 & 3.22E-02 & 17.33$\pm$0.01 & 0.59$\pm$0.01 & 0.31$\pm$0.02 & 20.81$\pm$0.15\tabularnewline
SGOBI & 182.21 & 1324.99$\pm$84.07 & 6.40E-02 & \textbf{5.46$\pm$0.00} & 0.48$\pm$0.02 & 0.41$\pm$0.03 & 14.41$\pm$0.08\tabularnewline
SGOBI{*} & 179.09 & 1291.31$\pm$96.20 & 4.99E-02 & 10.86$\pm$0.00 & 0.46$\pm$0.01 & 0.35$\pm$0.01 & 12.73$\pm$0.05\tabularnewline
\midrule 
GOSH & 179.10 & 1275.80$\pm$89.89 & 3.09E-02 & 7.98$\pm$0.01 & 0.42$\pm$0.01 & \textbf{0.19$\pm$0.01} & 8.31$\pm$0.07\tabularnewline
GOSH{*} & \textbf{176.66} & \textbf{1202.39$\pm$77.55} & \textbf{2.37E-02} & 15.98$\pm$0.01 & 0.51$\pm$0.02 & 0.33$\pm$0.02 & 12.23$\pm$0.09\tabularnewline
\bottomrule 
\end{tabular}
\label{tab:simulator}
\end{table*}

\subsection{Model Training and Assumptions}

For the GOSH and GOSH* algorithms, we use standard NPN models with the following characteristics adapted from \cite{tuli2021cosco}. We use the following non-affine activation functions for our neural approximators to be differentiable for all input values. Consider the number of hosts as $N$ and the maximum number of tasks as $M$.
\begin{enumerate}
    \item Input layer of size $|S_t| + M \times N$ for GOSH and $|S_t| + M \times N + 1$ for GOSH*. The non-linearity used here is $\textsf{softplus}$\footnote{The definitions of these activation functions can be seen at the PyTorch web-page: \url{https://pytorch.org/docs/stable/nn.html}.} as in \cite{tuli2020dynamic}. Note, $N$ may vary from 10 in tests in the physical environment to 50 in the simulator.
    \item FC/NPN layer of size $128$ with $\textsf{softplus}$ activation.
    \item FC/NPN layer of size $64$ with $\textsf{tanhshrink}$ activation.
    \item FC/NPN layer of size $1$ with $\textsf{sigmoid}$ activation.
\end{enumerate}
The $f$ model has NPN layers with same sizes as above. The $g$ and $h$ models had FC layers with the $g$ model using $50\%$ dropout for Bayesian inference as in prior work~\cite{wang2016natural}.

We generate training data for GOSH by running a random placement scheduler for $2000$ scheduling intervals to capture diverse state-space. For GOSH*, we create a separate dataset by running a random scheduler with pre-trained models of the GOSH scheduler to include $\bar{\mathcal{O}}_t$ in the datapoints. To implement and test GOSH and GOSH* models, we use the COSCO framework implemented in Python~\cite{tuli2021cosco}. The gradient calculation and optimization (Equation~\ref{eq:gosh}) are performed using the Autograd package~\cite{paszke2017automatic}. At training time, we use 50\% dropout in the teacher network and learning rate $10^{-5}$ {in the AdamW optimizer}. {We perform five-fold cross-validation for statistical significance, where we randomly sample 80\% of data to get the training set and the rest as the cross-validation set. All models are trained using the PyTorch 1.8 library. To prevent model overfitting, we use decoupled adaptive weight regularization from~\cite{loshchilov2017decoupled} using the weight decay parameter of $10^{-4}$. The values of the learning rate and the regularization hyperparameters are obtained via grid search using the \texttt{Ray Tune} API in PyTorch~\cite{raytune}. We also use the early stopping criterion to train the models~\cite{prechelt1998early}. The hyperparameter search is performed on the mean five-fold cross-validation accuracy.}

\subsection{Baselines and Evaluation Metrics}
\label{sec:baseslines}

We evaluate the performance of the GOSH/GOSH* model against the state-of-the-art approaches discussed in Section~\ref{sec:related_work}. These include the pessimistic-optimistic UCB optimization approach \textit{POND}~\cite{liu2020pond}, determenistic surrogate based methods \textit{GOBI} and \textit{GOBI*}~\cite{tuli2021cosco}, \textit{A3C}~\cite{tuli2020dynamic} and the \textit{HGP} model, where we use the LBFGS method with a HGP surrogate model for optimization as done in~\cite{jamshidi2016uncertainty, bui2017energy}.

We also compare the proposed methods with ablated models, where we replace either one of second-order optimization or heteroscedastic modeling components with simpler versions as described below.
\begin{itemize}
    \item HGOBI/HGOBI*: GOSH with first-order optimization (as in GOBI) instead of second-order. Thus, $LCB_t$ was optimized using first-order gradient descent. HGOBI* is the co-simulator aided extension of HGOBI.
    \item SGOBI/SGOBI*: GOSH with updating the output of an FCN neural model (as in GOBI). Thus, the optimization was done as per~\eqref{eq:sgobi}. SGOBI* is the co-simulator aided extension of SGOBI.
\end{itemize}

Comparison with the ablated models allows us to study individual contributions of both components.

\textbf{Evaluation Metrics:}
We use the following evaluation metrics in our experiments motivated from prior works~\cite{tuli2020ithermofog, tuli2021cosco, basu2019learn}. 
\begin{enumerate}
    \item \textit{Energy}:  To measure the energy consumption, we use the power consumption models taken from the commonly-used SPEC benchmarks repository~\cite{spec}.
    \item \textit{SLA Violations}: A task's SLA is violated if its response time is greater than the deadline. We consider the relative definition of SLA (as in~\cite{tuli2021cosco}) where the deadline is the 95$^{th}$ percentile response time \blue{for the same application type} on the state of the art baseline \textit{GOBI*}. To ensure sufficient sample size, we run the \textit{GOBI*} baseline for 1000 scheduling intervals and evaluate the 95$^{th}$ percentile response times for each application. \blue{We define the \textit{fraction of SLA violations for each application} as the ratio of the application instances where SLA was violated to the total number of instances running in a trace of this type. Also, \textit{fraction of SLA violations for a scheduler} is defined as the ratio of the application instances where SLA was violated to the total instances run in a trace across all application types.}
    \item \textit{Fairness}: We use the Jain's fairness index over the IPS of the running tasks in the system. 
    \item \textit{Average Migration Time} which is the average time for task (container) migrations.
\end{enumerate}
We also compare the \textit{Response Time}, \textit{Scheduling Time} and \textit{Average Wait Time} of all executed tasks. \blue{For a bag-of-task workload model, each workload container is composed of multiple tasks (inference over different data inputs). Each task is independent of the other task within the same running application instance, allowing parallel execution. However, when an application instance, realized as a Docker container, is scheduled on a host with limited available resources, the tasks within the application may be run sequentially. We define the \textit{average response time of a scheduler} as the mean response time of all completed tasks in an execution trace using the particular scheduler for task placement and migration. We define the \textit{average response time per application} as the time it takes to execute all running tasks within the same application instance. This means the timestamp from the start of execution of the first task to the timestamp when the last task completes execution.}

\begin{figure*}[t]
    \centering
    \subfigure[Uncertainties with Interval]{
    \includegraphics[width=.78\columnwidth]{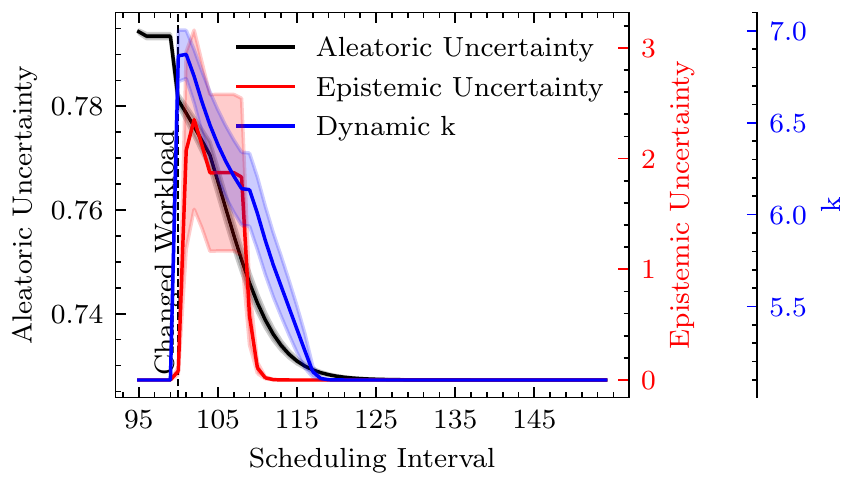}
    \label{fig:sensitivity_k_uncertainty}
    }
    \subfigure[Comparison of the adaptability with $k$]{
    \includegraphics[width=.66\columnwidth]{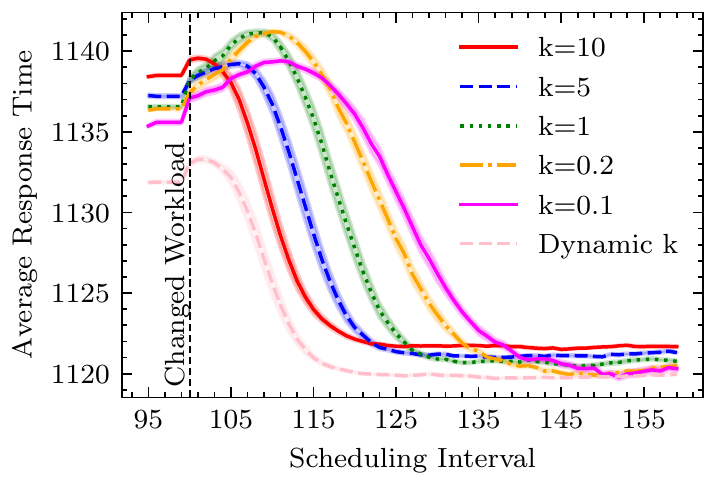}
    \label{fig:sensitivity_k_line}
    }
    \subfigure[Average Response Time]{
    \includegraphics[width=.52\columnwidth]{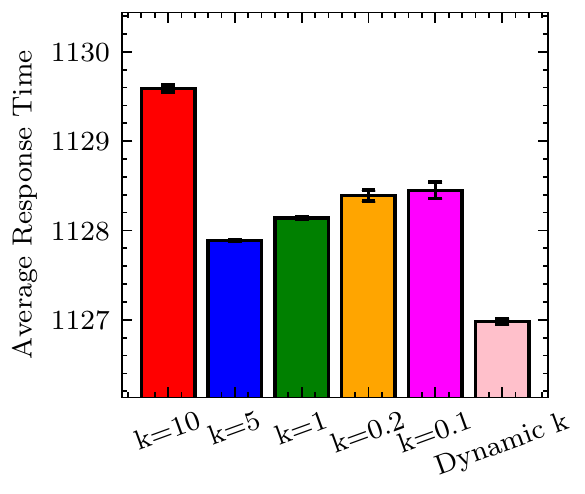}
    \label{fig:sensitivity_k_rt}
    }
    \caption{Sensitivity Analysis for $k$} \vspace{-1.2em}
    \label{fig:sensitivity}
\end{figure*}

\subsection{Results and Ablation Analysis}

We now provide comparative results showing the performance of the GOSH and GOSH* schedulers against the baselines and ablated models (described in Section~\ref{sec:baseslines}). Figure~\ref{fig:framework_results} shows results for the physical setup with 10 Azure hosts and Table~\ref{tab:simulator} for the simulation platform with 50 hosts.

Figure~\ref{fig:f_energy} shows the average energy consumption per task for the Azure hosts, with Figure~\ref{fig:f_energy_series} showing energy consumption with execution time. GOSH and GOSH* (\blue{$150.9 - 157.6$} KW-hr respectively) have lower energy consumption than the GOBI* approach by \blue{$17.91\%-21.40\%$} where GOBI* has the least consumption among the baselines. This is because of the ability of the second-order optimization approach to reach better objective scores and hence reduce the $AEC_t$ metric, as the SGOBI/SGOBI* models also reach low energy consumption values. Further, the reduced average response time for GOSH and GOSH* models gives a higher number of completed tasks (Figure~\ref{fig:f_containers_series}) and hence lower average energy consumption. A similar trend is also seen for the simulation setup (Table~\ref{tab:simulator}) where GOBI* achieves the least energy consumption among the baselines (\blue{$211.05$} KW-hr) and GOSH/GOSH* improve upon this by \blue{$15.13\% - 16.29\%$}.

Figure~\ref{fig:f_response} compares the average response time for each scheduler. Again, the GOBI* scheduler has the least average response time of \blue{$154\pm29$} seconds. Here, GOSH* has much lower response times (\blue{$28.57\%$} lower than GOBI*) and the GOSH scheduler has \blue{$28.29\%$} lower than the GOBI model. Again, this is because of the second-order optimization, which gives low scheduling overhead and can reach better objective scores of $ART_t$. \blue{We see here that the ablated model HGOBI* has higher response times than the GOBI* model. Specifically, HGOBI* gives $16.88\%$ higher response time than GOBI*. On the contrary, the HGOBI* model has} a significantly lower SLA violation rate compared to the best baseline GOBI* (relative improvement is \blue{$44.21\%$}) as seen in Figure~\ref{fig:f_sla}. This is because the HGOBI* model aims to minimize the value-at-risk instead of the mean response time, reducing the chance of violating the SLA deadline. Among the baselines, the GOBI* model has the lowest SLA violation rate (\blue{$5.82\%$}) with the HGP model coming second (\blue{$8.21\%$}). The GOSH and GOSH* models beat the GOBI* scheduler giving SLA violation rates of just \blue{$3.04\% - 1.11\%$}. Similar trends are seen in the case of 50 simulated hosts where the GOSH and GOSH* models give \blue{$14.25\% - 19.16\%$} lower average response time and \blue{$55.98\% - 66.23\%$}  lower SLA violation rate.

For the three applications, the average response time and SLA violation for each scheduler are shown in Figures~\ref{fig:f_response_pa} and \ref{fig:f_sla_pa}. The high response time of the POND baseline is mostly attributed to frequent migrations performed by this method (see Figure~\ref{fig:f_migration_time}). On the other hand, GOSH* is the most parsimonious strategy in task migrations, allowing lower response times than other approaches. The deviations from the $5\%$ mark in Figure~\ref{fig:f_sla_pa} is due to the stochasticity in the workloads, motivating us to fine-tune the model at each interval (line~8 in Alg.~\ref{alg:gosh2}). Clearly, among the DeFog benchmark applications, Yolo takes the maximum time. {As seen previously, Figure~\ref{fig:f_sla_pa} shows the SLA violations rates for each application. The SLA violation rates of the SGOBI and SGOBI* methods are higher than the heteroscedastic models GOSH, GOSH*, HGOBI and HGOBI*. Even the HGP baseline has a lower SLA violation rate than the SGOBI model for the Yolo application. This shows the importance of having heteroscedastic modeling in estimating QoS scores. These improvements are at the cost of task execution fairness.}

{In terms of the scheduling time, as seen in Figures~\ref{fig:f_scheduling_time} and~\ref{fig:f_scheduling_time_series}, the co-simulation based variant of GOBI, HGOBI, SGOBI and GOSH have higher scheduling times. This is due to the additional step that calls the co-simulator to estimate the QoS scores of the next interval (see line 6 in Alg.~\ref{alg:gosh2}). GOBI has the least scheduling time among baselines, showing the importance of gradient-based optimization. However, the second-order optimization further reduces the scheduling time of SGOBI and SGOBI* by $11.79\% - 20.90\%$ compared to GOBI and GOBI* respectively in the large-scale setting of 50 hosts (see Table~\ref{tab:simulator}). However, having a more sophisticated neural network (NPN) in HGOBI and HGOBI* increases the scheduling time compared to GOBI and GOBI* due to a slight increase in the time taken to fine-tune the model (line~31 in Alg.~\ref{alg:gosh} and line~8 in Alg.~\ref{alg:gosh2}). A more detailed comparison of the model overheads is given in Section~\ref{sec:overhead}. Figure~\ref{fig:f_fairness} shows that the GOBI* baseline is the fairest among all models with a Jain's fairness index of \blue{0.75}. GOSH* has the second highest fairness of \blue{0.57}, higher than all baselines other than GOBI*.} The average wait times for applications with execution is shown in Figure~\ref{fig:f_wait_time_series}. A3C has high wait times due to the inability of the RL baseline to quickly adapt in cases of hosts running and capacity.  Comparing other metrics like fairness, GOSH and GOSH* have lower Jain's fairness index value than the GOBI and GOBI* models. Among all models, the GOBI* model has the highest fairness index, with GOSH and GOSH* having only $0.26$ and \blue{$0.57$} respectively. The scheduling time is lowest for the SGOBI ablated model of $2.33$ seconds in the physical setup and $5.46$ seconds in the simulated setup. Moreover, the scheduling overhead of GOSH compared to GOBI is $12.72\%$ and that of GOSH* compared to GOBI* is $15.22\%$. Similar overheads are observed for the 50 simulated hosts case of $16.39\% - 28.9\%$.

\subsection{Additional Experiments}
\label{sec:sensitivity}
\label{sec:agility}
\label{sec:overhead}

\begin{table}[]
    \centering
    \caption{Loss values of various models on test data. {The values are averaged over the five-fold cross-validation runs.}}
    \begin{tabular}{@{}lccc@{}}
    \toprule 
    \textbf{Scheduler} & \textbf{Model} & \textbf{MSE} & \textbf{KLD}\tabularnewline
    \midrule
    GOBI & FCN & 0.2582 & 2.3344\tabularnewline
     
    GOBI{*} & FCN & \textbf{0.1176} & 1.9837\tabularnewline
     
    HGP & HGP & 0.3821 & 1.0928\tabularnewline
     
    GOSH & NPN & 0.2812 & 0.9236\tabularnewline
     
    GOSH{*} & NPN & 0.1182 & \textbf{0.3542}\tabularnewline
    \bottomrule 
    \end{tabular}
    \label{tab:loss}
\end{table}

\textbf{Modelling Accuracy:}
Experiments on the dataset collected using a random scheduler show that the NPN model is able to give lower KL divergence between the predicted output and the ground-truth compared to the baseline FCN model, even though the Mean-Square Error is slightly higher ($0.5-8.9\%$, see Table~\ref{tab:loss}). The NPN models perform even better than the HGP models used in prior work~\cite{jamshidi2016uncertainty, bui2017energy}, giving more accurate predictions and hence improving QoS.

\textbf{Sensitivity Analysis:}
We now show the importance of dynamically updating the value of the $k$ parameter based on the epistemic uncertainty. We conduct our experiments on the simulated setup with the Bitbrain workload traces. We run the GOSH model for 200 intervals with static values of $k$ and one with $k=5$ initialization being updated at each interval as per~\eqref{eq:dynamic_k}. We keep the workloads as Bitbrain random traces for the first 100 intervals and the sequential traces for the next 100 intervals. The "Changed Workload" vertical line in Figure~\ref{fig:sensitivity} highlights the instant of this workload transition. {We average our results over 20 runs for 5 cases of static $k$ with the dynamic $k$ setup.} 

As is apparent from Figure~\ref{fig:sensitivity_k_uncertainty}, the aleatoric uncertainty in the random traces is higher than the sequential traces. The sharp fall in the epistemic uncertainty (Figure~\ref{fig:sensitivity_k_uncertainty}) is due to the fact that the model explores and quickly reaches a local region of "well known" state space. Figure~\ref{fig:sensitivity_k_line} shows the average response time per interval of leaving tasks. The converged response time (before the 100$^{th}$ and after the 150$^{th}$ interval) is lower for low values of $k$ as a higher relative weight is given to the value-at-risk metric. However, as $k$ increases, the model takes more time to adapt to the new scenario, as shown by the slopes in Figure~\ref{fig:sensitivity_k_line}, indicating that higher $k$ enforces higher exploration and swift adaptability. {Finally, Figure~\ref{fig:sensitivity_k_rt} shows that the average response time over the complete 200 interval duration is also minimum for the dynamic $k$ setup. This indicates that dynamically adapting the $k$ parameter is crucial for quick adaptability in volatile environments.}

\begin{figure}[t]
    \centering
    \includegraphics[width=\columnwidth]{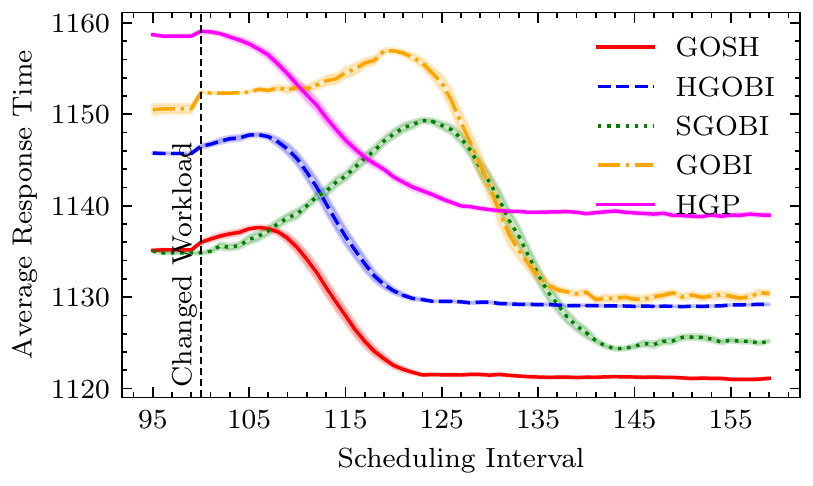}
    \caption{Comparison of the adaptability of various models.}
    \label{fig:adaptability}
\end{figure}

\textbf{Agility Testing:}
Figure~\ref{fig:adaptability} compares how quickly GOSH is able to adapt to the new setting compared to baselines. As we see, GOSH, HGOBI and HGP are able to adapt quickly to new settings compared to GOBI or SGOBI. Both SGOBI and GOBI take twice the time (40 intervals) to converge to the new scenario compared to GOSH. Thus, heteroscedastic modeling aids quick adaptability in the proposed methods.

\begin{figure}[t]
    \centering
    \includegraphics[width=\columnwidth]{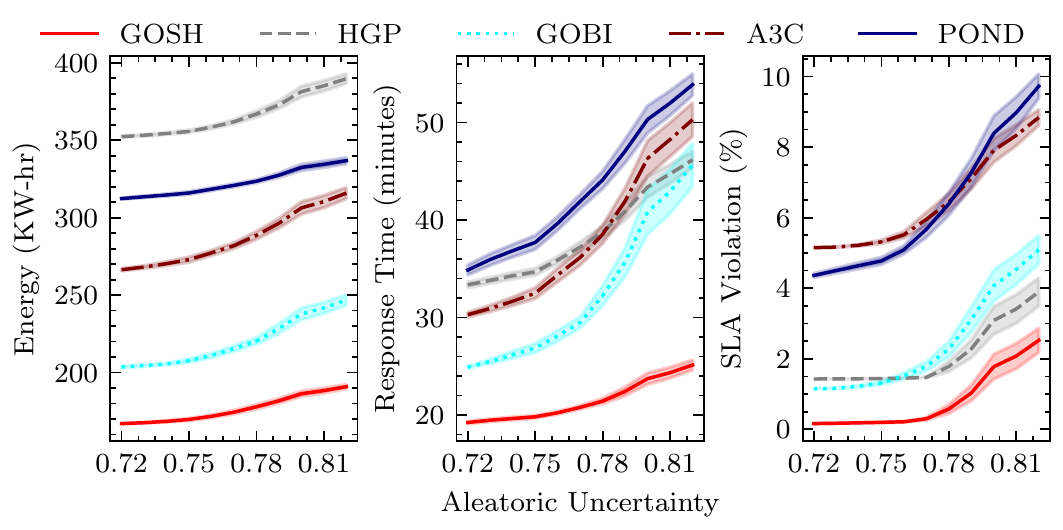}
    \caption{Comparison of the performance with increasing uncertainty.}
    \label{fig:perf_uncertainty}
\end{figure}
\begin{figure}[t]
    \centering
    \includegraphics[width=\columnwidth]{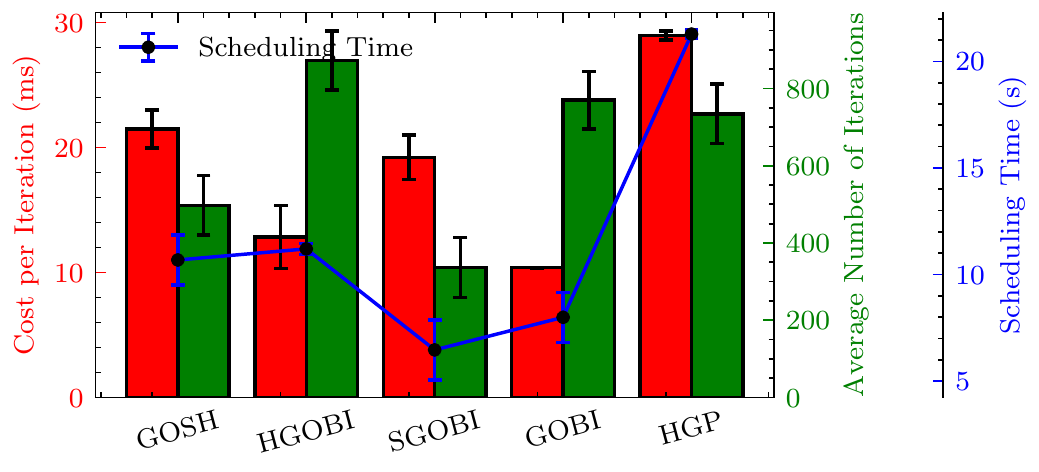}
    \caption{Comparison of the scheduling overhead of various models.}
    \label{fig:overhead}
\end{figure}

\textbf{Performance with Uncertainty:} Figure~\ref{fig:perf_uncertainty} compares the performance of the GOSH approach with the baseline models as the aleatoric uncertainty of the system increases. We exclude the co-simulated extensions to focus on the effects of second-order optimization and heteroscedastic modeling. To test this, we vary the ratio of the \textit{sequential} and \textit{random} traces of the Bitbrain dataset (with the former having lower uncertainty than the latter). We run them sequentially with the \textit{sequential} only workloads first. The \textit{sequential} only workload traces give the lower bound (0.72) and \textit{random} only traces give the upper bound (0.81) of the uncertainty considered across the $x$ axis in Figure~\ref{fig:perf_uncertainty}. In terms of energy consumption, response time and SLA violation rate, GOSH not only has better metric scores, but also does not face excessive performance drop in highly uncertain cases, thanks to its gradient regularization.  

\textbf{Overhead Comparison:} Figure~\ref{fig:overhead} compares the cost per iteration (in milliseconds), average number of iterations and the overall scheduling time for GOSH and baseline models. The per-iteration cost of SGOBI is higher due to the additional step of Hutchinson's method and moving average computation. However, the improvement due to the reduction in the number of iterations is more than this excess cost, so much so that the overall scheduling time of SGOBI is lower than GOBI.
This allows the overall overhead of GOSH due to the heteroscedastic modeling to not be as high as GOBI ($33.65\%$) compared to the HGOBI approach ($40.12\%$).

\section{Conclusions and Future Work}
\label{sec:conclusions}

In this work, we have presented multiple contributions including heteroscedastic modeling for aleatoric uncertainty estimation and value-at-risk calculation, a teacher-student model for epistemic uncertainty estimation, efficient second-order optimization and feedback based exploration control.  All these contributions allow GOSH and GOSH* to out-perform the baseline models in terms of QoS metrics such as energy, response time and SLA violation by up to $18\%$, $27\%$ and $82\%$ respectively in a heterogeneous fog environment with real-world workloads. As shown in Figure~\ref{fig:f_scheduling_time}, between GOSH and GOSH*, the former is more suitable for environments with a resource-constrained fog broker, whereas the latter reaches a better objective score but requires heavier computational resources.

For the future, we propose the following extensions. The GOSH and GOSH schedulers could be adapted for serverless frameworks allowing us fine-grained orchestration with increased productivity and flexibility, reducing the container management overheads. Further, other stochastic surrogate models like distributed Bayesian Neural Networks may be explored for decentralized scheduling with higher modeling performance and lower resource consumption overheads per host.

\section*{Software Availability}
The code is available at \url{https://github.com/imperial-qore/COSCO/tree/gosh}. The Docker images used in the experiments are available at \url{https://hub.docker.com/u/shreshthtuli}. The training datasets are available at \url{https://doi.org/10.5281/zenodo.4897944}, released under the CC BY 4.0 license.

\section*{Acknowledgments}
Shreshth Tuli is supported by the President's PhD scholarship at Imperial College London.

\bibliographystyle{IEEEtran}
\bibliography{references}

\begin{thebibliography}{10}
\providecommand{\url}[1]{#1}
\csname url@samestyle\endcsname
\providecommand{\newblock}{\relax}
\providecommand{\bibinfo}[2]{#2}
\providecommand{\BIBentrySTDinterwordspacing}{\spaceskip=0pt\relax}
\providecommand{\BIBentryALTinterwordstretchfactor}{4}
\providecommand{\BIBentryALTinterwordspacing}{\spaceskip=\fontdimen2\font plus
\BIBentryALTinterwordstretchfactor\fontdimen3\font minus
  \fontdimen4\font\relax}
\providecommand{\BIBforeignlanguage}[2]{{%
\expandafter\ifx\csname l@#1\endcsname\relax
\typeout{** WARNING: IEEEtran.bst: No hyphenation pattern has been}%
\typeout{** loaded for the language `#1'. Using the pattern for}%
\typeout{** the default language instead.}%
\else
\language=\csname l@#1\endcsname
\fi
#2}}
\providecommand{\BIBdecl}{\relax}
\BIBdecl

\bibitem{yousefpour2019all}
A.~Yousefpour, C.~Fung, T.~Nguyen, K.~Kadiyala, F.~Jalali, A.~Niakanlahiji,
  J.~Kong, and J.~P. Jue, ``All one needs to know about fog computing and
  related edge computing paradigms: A complete survey,'' \emph{Journal of
  Systems Architecture}, vol.~98, pp. 289--330, 2019.

\bibitem{tuli2019fogbus}
S.~Tuli, R.~Mahmud, S.~Tuli, and R.~Buyya, ``Fogbus: A blockchain-based
  lightweight framework for edge and fog computing,'' \emph{Journal of Systems
  and Software}, vol. 154, pp. 22--36, 2019.

\bibitem{tuli2021cosco}
S.~Tuli, S.~Poojara, S.~N. Srirama, G.~Casale, and N.~Jennings, ``{COSCO:
  Container Orchestration using Co-Simulation and Gradient Based Optimization
  for Fog Computing Environments},'' \emph{IEEE Transactions on Parallel and
  Distributed Systems}, 2021.

\bibitem{liang2020ai}
Q.~Liang, P.~Shenoy, and D.~Irwin, ``{AI on the edge: Characterizing AI-based
  IoT applications using specialized edge architectures},'' in \emph{2020 IEEE
  International Symposium on Workload Characterization (IISWC)}.\hskip 1em plus
  0.5em minus 0.4em\relax IEEE, 2020, pp. 145--156.

\bibitem{tuli2021generative}
S.~Tuli, S.~Tuli, G.~Casale, and N.~R. Jennings, ``Generative optimization
  networks for memory efficient data generation,'' \emph{Advances in Neural
  Information Processing Systems, Workshop on ML for Systems}, 2021.

\bibitem{chang2019internet}
C.~Chang, S.~N. Srirama, and R.~Buyya, ``{Internet of Things (IoT) and new
  computing paradigms},'' \emph{Fog and edge computing: principles and
  paradigms}, vol.~6, pp. 1--23, 2019.

\bibitem{matrouk2021scheduling}
K.~Matrouk and K.~Alatoun, ``Scheduling algorithms in fog computing: A
  survey,'' \emph{International Journal of Networked and Distributed
  Computing}, 2021.

\bibitem{basu2019learn}
D.~Basu, X.~Wang, Y.~Hong, H.~Chen, and S.~Bressan, ``Learn-as-you-go with
  megh: Efficient live migration of virtual machines,'' \emph{IEEE Transactions
  on Parallel and Distributed Systems}, vol.~30, no.~8, pp. 1786--1801, 2019.

\bibitem{tuli2020dynamic}
S.~Tuli, S.~Ilager, K.~Ramamohanarao, and R.~Buyya, ``{Dynamic Scheduling for
  Stochastic Edge-Cloud Computing Environments using A3C learning and Residual
  Recurrent Neural Networks},'' \emph{IEEE Transactions on Mobile Computing},
  2020.

\bibitem{ghosal2020deep}
G.~R. Ghosal, D.~Ghosal, A.~Sim, A.~V. Thakur, and K.~Wu, ``A deep
  deterministic policy gradient based network scheduler for deadline-driven
  data transfers,'' in \emph{2020 IFIP Networking Conference
  (Networking)}.\hskip 1em plus 0.5em minus 0.4em\relax IEEE, 2020, pp.
  253--261.

\bibitem{liu2020pond}
X.~Liu, B.~Li, P.~Shi, and L.~Ying, ``{POND: Pessimistic-Optimistic oNline
  Dispatch},'' \emph{arXiv preprint arXiv:2010.09995}, 2020.

\bibitem{panda2015uncertainty}
S.~Panda, B.~Neha, and S.~Sathua, ``An uncertainty-based task scheduling for
  heterogeneous multi-cloud systems,'' \emph{International Journal of
  Information Processing}, vol.~9, no.~2, pp. 13--24, 2015.

\bibitem{tychogiorgos2013non}
G.~Tychogiorgos, A.~Gkelias, and K.~K. Leung, ``A non-convex distributed
  optimization framework and its application to wireless ad-hoc networks,''
  \emph{IEEE Transactions on Wireless Communications}, vol.~12, no.~9, pp.
  4286--4296, 2013.

\bibitem{zhang2018double}
Q.~Zhang, M.~Lin, L.~T. Yang, Z.~Chen, S.~U. Khan, and P.~Li, ``A double deep
  q-learning model for energy-efficient edge scheduling,'' \emph{IEEE
  Transactions on Services Computing}, vol.~12, no.~5, pp. 739--749, 2018.

\bibitem{gazori2019saving}
P.~Gazori, D.~Rahbari, and M.~Nickray, ``Saving time and cost on the scheduling
  of fog-based iot applications using deep reinforcement learning approach,''
  \emph{Future Generation Computer Systems}, 2019.

\bibitem{bae2019beyond}
J.~Bae, J.~Lee, and S.~Chong, ``Beyond max-weight scheduling: A reinforcement
  learning-based approach,'' in \emph{2019 International Symposium on Modeling
  and Optimization in Mobile, Ad Hoc, and Wireless Networks (WiOPT)}.\hskip 1em
  plus 0.5em minus 0.4em\relax IEEE, 2019, pp. 1--8.

\bibitem{krishnasamy2018augmenting}
S.~Krishnasamy, P.~Akhil, A.~Arapostathis, R.~Sundaresan, and S.~Shakkottai,
  ``Augmenting max-weight with explicit learning for wireless scheduling with
  switching costs,'' \emph{IEEE/ACM Transactions on Networking}, vol.~26,
  no.~6, pp. 2501--2514, 2018.

\bibitem{gruian2003uncertainty}
F.~Gruian and K.~Kuchcinski, ``Uncertainty-based scheduling: energy-efficient
  ordering for tasks with variable execution time,'' in \emph{Proceedings of
  the 2003 international symposium on Low power electronics and design}, 2003,
  pp. 465--468.

\bibitem{zinnen2011deadline}
A.~Zinnen and T.~Engel, ``Deadline constrained scheduling in hybrid clouds with
  gaussian processes,'' in \emph{2011 International Conference on High
  Performance Computing \& Simulation}.\hskip 1em plus 0.5em minus 0.4em\relax
  IEEE, 2011, pp. 294--300.

\bibitem{jamshidi2016uncertainty}
P.~Jamshidi and G.~Casale, ``An uncertainty-aware approach to optimal
  configuration of stream processing systems,'' in \emph{2016 IEEE 24th
  International Symposium on Modeling, Analysis and Simulation of Computer and
  Telecommunication Systems (MASCOTS)}.\hskip 1em plus 0.5em minus 0.4em\relax
  IEEE, 2016, pp. 39--48.

\bibitem{bui2017energy}
D.-M. Bui, Y.~Yoon, E.-N. Huh, S.~Jun, and S.~Lee, ``Energy efficiency for
  cloud computing system based on predictive optimization,'' \emph{Journal of
  Parallel and Distributed Computing}, vol. 102, pp. 103--114, 2017.

\bibitem{tchernykh2015towards}
A.~Tchernykh, U.~Schwiegelsohn, V.~Alexandrov, and E.-g. Talbi, ``Towards
  understanding uncertainty in cloud computing resource provisioning,''
  \emph{Procedia Computer Science}, vol.~51, pp. 1772--1781, 2015.

\bibitem{tuli2021pregan}
S.~Tuli, G.~Casale, and N.~R. Jennings, ``{PreGAN: Preemptive Migration
  Prediction Network for Proactive Fault-Tolerant Edge Computing},'' in
  \emph{IEEE Conference on Computer Communications (INFOCOM)}.\hskip 1em plus
  0.5em minus 0.4em\relax IEEE, 2022.

\bibitem{kochenderfer2019algorithms}
M.~J. Kochenderfer and T.~A. Wheeler, \emph{Algorithms for optimization}.\hskip
  1em plus 0.5em minus 0.4em\relax MIT Press, 2019.

\bibitem{aima}
S.~Russell and P.~Norvig, \emph{Artificial Intelligence: A Modern Approach},
  3rd~ed.\hskip 1em plus 0.5em minus 0.4em\relax USA: Prentice Hall Press,
  2009.

\bibitem{nandi2001artificial}
S.~Nandi, S.~Ghosh, S.~S. Tambe, and B.~D. Kulkarni, ``Artificial
  neural-network-assisted stochastic process optimization strategies,''
  \emph{AIChE journal}, vol.~47, no.~1, pp. 126--141, 2001.

\bibitem{jawad2018robust}
M.~Jawad, M.~B. Qureshi, U.~Khan, S.~M. Ali, A.~Mehmood, B.~Khan, X.~Wang, and
  S.~U. Khan, ``A robust optimization technique for energy cost minimization of
  cloud data centers,'' \emph{IEEE Transactions on Cloud Computing}, 2018.

\bibitem{da2018resource}
R.~A. da~Silva and N.~L. da~Fonseca, ``Resource allocation mechanism for a
  fog-cloud infrastructure,'' in \emph{2018 IEEE International Conference on
  Communications (ICC)}.\hskip 1em plus 0.5em minus 0.4em\relax IEEE, 2018, pp.
  1--6.

\bibitem{tuli2021mcds}
S.~Tuli, G.~Casale, and N.~R. Jennings, ``{MCDS: AI Augmented Workflow
  Scheduling in Mobile Edge Cloud Computing Systems},'' \emph{arXiv preprint
  arXiv:2112.07269}, 2021.

\bibitem{adamw}
I.~Loshchilov and F.~Hutter, ``Decoupled weight decay regularization,'' in
  \emph{International Conference on Learning Representations}, 2018.

\bibitem{pan2015annealed}
H.~Pan and H.~Jiang, ``Annealed gradient descent for deep learning,'' in
  \emph{The Thirty-First Conference on Uncertainty in Artificial Intelligence},
  2015, pp. 652--661.

\bibitem{loshchilov2016sgdr}
I.~Loshchilov and F.~Hutter, ``{SGDR: Stochastic gradient descent with warm
  restarts},'' in \emph{International Conference on Learning Representations
  (ICLR)}, 2017.

\bibitem{shaker2020aleatoric}
M.~H. Shaker and E.~H{\"u}llermeier, ``Aleatoric and epistemic uncertainty with
  random forests,'' in \emph{International Symposium on Intelligent Data
  Analysis}.\hskip 1em plus 0.5em minus 0.4em\relax Springer, 2020, pp.
  444--456.

\bibitem{hochreiter1997long}
S.~Hochreiter and J.~Schmidhuber, ``Long short-term memory,'' \emph{Neural
  computation}, vol.~9, no.~8, pp. 1735--1780, 1997.

\bibitem{lecun2007energy}
Y.~LeCun, S.~Chopra, M.~Ranzato, and F.-J. Huang, ``Energy-based models in
  document recognition and computer vision,'' in \emph{Ninth International
  Conference on Document Analysis and Recognition (ICDAR 2007)}, vol.~1.\hskip
  1em plus 0.5em minus 0.4em\relax IEEE, 2007, pp. 337--341.

\bibitem{wang2016natural}
H.~Wang, X.~Shi, and D.-Y. Yeung, ``Natural-parameter networks: a class of
  probabilistic neural networks,'' in \emph{Proceedings of the 30th
  International Conference on Neural Information Processing Systems}, 2016, pp.
  118--126.

\bibitem{rezaei2020mean}
F.~Rezaei, A.~A. Najafi, and R.~Ramezanian, ``Mean-conditional value at risk
  model for the stochastic project scheduling problem,'' \emph{Computers \&
  Industrial Engineering}, vol. 142, p. 106356, 2020.

\bibitem{liu2020improving}
Y.~Liu, B.~Cao, and H.~Li, ``Improving ant colony optimization algorithm with
  epsilon greedy and levy flight,'' \emph{Complex \& Intelligent Systems}, pp.
  1--12, 2020.

\bibitem{sahlin2021we}
U.~Sahlin, I.~Helle, and D.~Perepolkin, ``“this is what we don't know”:
  Treating epistemic uncertainty in bayesian networks for risk assessment,''
  \emph{Integrated Environmental Assessment and Management}, vol.~17, no.~1,
  pp. 221--232, 2021.

\bibitem{postels2019sampling}
J.~Postels, F.~Ferroni, H.~Coskun, N.~Navab, and F.~Tombari, ``Sampling-free
  epistemic uncertainty estimation using approximated variance propagation,''
  in \emph{Proceedings of the IEEE/CVF International Conference on Computer
  Vision}, 2019, pp. 2931--2940.

\bibitem{matiisen2019teacher}
T.~Matiisen, A.~Oliver, T.~Cohen, and J.~Schulman, ``Teacher--student
  curriculum learning,'' \emph{IEEE transactions on neural networks and
  learning systems}, vol.~31, no.~9, pp. 3732--3740, 2019.

\bibitem{gal2016dropout}
Y.~Gal and Z.~Ghahramani, ``Dropout as a bayesian approximation: Representing
  model uncertainty in deep learning,'' in \emph{international conference on
  machine learning}.\hskip 1em plus 0.5em minus 0.4em\relax PMLR, 2016, pp.
  1050--1059.

\bibitem{kendall2017uncertainties}
A.~Kendall and Y.~Gal, ``What uncertainties do we need in bayesian deep
  learning for computer vision?'' in \emph{Proceedings of the 31st
  International Conference on Neural Information Processing Systems}, 2017, pp.
  5580--5590.

\bibitem{yao2021adahessian}
Z.~Yao, A.~Gholami, S.~Shen, M.~Mustafa, K.~Keutzer, and M.~Mahoney,
  ``Adahessian: An adaptive second order optimizer for machine learning,'' in
  \emph{Proceedings of the AAAI Conference on Artificial Intelligence},
  vol.~35, no.~12, 2021, pp. 10\,665--10\,673.

\bibitem{martens2015optimizing}
J.~Martens and R.~Grosse, ``Optimizing neural networks with kronecker-factored
  approximate curvature,'' in \emph{International conference on machine
  learning}.\hskip 1em plus 0.5em minus 0.4em\relax PMLR, 2015, pp. 2408--2417.

\bibitem{yao2018hessian}
Z.~Yao, A.~Gholami, K.~Keutzer, and M.~W. Mahoney, ``Hessian-based analysis of
  large batch training and robustness to adversaries,'' in \emph{Proceedings of
  the 32nd International Conference on Neural Information Processing Systems},
  2018, pp. 4954--4964.

\bibitem{yao2020pyhessian}
------, ``Pyhessian: Neural networks through the lens of the hessian,'' in
  \emph{2020 IEEE International Conference on Big Data (Big Data)}.\hskip 1em
  plus 0.5em minus 0.4em\relax IEEE, 2020, pp. 581--590.

\bibitem{loshchilov2017decoupled}
I.~Loshchilov and F.~Hutter, ``Decoupled weight decay regularization,''
  \emph{arXiv preprint arXiv:1711.05101}, 2017.

\bibitem{maroti2019rbed}
A.~Maroti, ``Rbed: Reward based epsilon decay,'' \emph{arXiv preprint
  arXiv:1910.13701}, 2019.

\bibitem{barrett2020implicit}
D.~Barrett and B.~Dherin, ``Implicit gradient regularization,'' in
  \emph{International Conference on Learning Representations}, 2020.

\bibitem{tuli2020healthfog}
S.~Tuli, N.~Basumatary, S.~S. Gill, M.~Kahani, R.~C. Arya, G.~S. Wander, and
  R.~Buyya, ``{HealthFog: An ensemble deep learning based smart healthcare
  system for automatic diagnosis of heart diseases in integrated IoT and fog
  computing environments},'' \emph{Future Generation Computer Systems}, vol.
  104, pp. 187--200, 2020.

\bibitem{tuli2021hunter}
S.~Tuli, S.~S. Gill, M.~Xu, P.~Garraghan, R.~Bahsoon, S.~Dustdar,
  R.~Sakellariou, O.~Rana, R.~Buyya, G.~Casale \emph{et~al.}, ``{HUNTER: AI
  based Holistic Resource Management for Sustainable Cloud Computing},''
  \emph{Journal of Systems and Software}, 2021.

\bibitem{ahmed2018docker}
A.~Ahmed and G.~Pierre, ``Docker container deployment in fog computing
  infrastructures,'' in \emph{2018 IEEE International Conference on Edge
  Computing (EDGE)}.\hskip 1em plus 0.5em minus 0.4em\relax IEEE, 2018, pp.
  1--8.

\bibitem{shen2015statisticalBitBrain}
S.~Shen, V.~van Beek, and A.~Iosup, ``Statistical characterization of
  business-critical workloads hosted in cloud datacenters,'' in \emph{15th
  IEEE/ACM International Symposium on Cluster, Cloud and Grid Computing}.\hskip
  1em plus 0.5em minus 0.4em\relax IEEE, 2015, pp. 465--474.

\bibitem{mcchesney2019defog}
J.~McChesney, N.~Wang, A.~Tanwer, E.~de~Lara, and B.~Varghese, ``Defog: fog
  computing benchmarks,'' in \emph{The 4th ACM/IEEE Symposium on Edge
  Computing}, 2019, pp. 47--58.

\bibitem{paszke2017automatic}
A.~Paszke, S.~Gross, S.~Chintala, G.~Chanan, E.~Yang, Z.~DeVito, Z.~Lin,
  A.~Desmaison, L.~Antiga, and A.~Lerer, ``Automatic differentiation in
  pytorch,'' 2017.

\bibitem{raytune}
``{PyTorch Ray Tune},''
  \url{https://pytorch.org/tutorials/beginner/hyperparameter_tuning_tutorial.html},
  accessed: 2021-10-20.

\bibitem{prechelt1998early}
L.~Prechelt, ``Early stopping-but when?'' in \emph{Neural Networks: Tricks of
  the trade}.\hskip 1em plus 0.5em minus 0.4em\relax Springer, 1998, pp.
  55--69.

\bibitem{tuli2020ithermofog}
S.~Tuli, S.~S. Gill, G.~Casale, and N.~R. Jennings, ``{iThermoFog: IoT-Fog
  based automatic thermal profile creation for cloud data centers using
  artificial intelligence techniques},'' \emph{Internet Technology Letters},
  vol.~3, no.~5, p. e198, 2020.

\bibitem{spec}
\BIBentryALTinterwordspacing
{Standard Performance Evaluation Corporation}. SPEC power consumption models.
  [Online]. Available: \url{https://www.spec.org/cloud\_iaas2018/results/}
\BIBentrySTDinterwordspacing

\end{thebibliography}

\end{document}